\documentclass[aip,amsmath,amssymb,reprint]{revtex4-1}
\usepackage{graphicx}
\usepackage{dcolumn}
\usepackage{bm}

\usepackage[utf8]{inputenc}
\usepackage[T1]{fontenc}
\usepackage{mathptmx}
\usepackage{etoolbox}

\usepackage{color}

\makeatletter
\def\@email#1#2{%
 \endgroup
 \patchcmd{\titleblock@produce}
  {\frontmatter@RRAPformat}
  {\frontmatter@RRAPformat{\produce@RRAP{*#1\href{mailto:#2}{#2}}}\frontmatter@RRAPformat}
  {}{}
}%

\newcommand*{\citen}[1]{%
	\begingroup
	\romannumeral-`\x 
	\setcitestyle{numbers}%
	\cite{#1}%
	\endgroup
}

\makeatother

\begin{document}
\title{Nonlinear bias of collective oscillation frequency induced by asymmetric Cauchy noise}

\author{Maria V.\ Ageeva}
\affiliation{Institute of Continuous Media Mechanics, UB RAS, Academician Korolev
Street 1, 614013 Perm, Russia}
\author{Denis S.\ Goldobin}
\affiliation{Institute of Continuous Media Mechanics, UB RAS, Academician Korolev
Street 1, 614013 Perm, Russia}
\affiliation{Institute of Physics and Mathematics, Perm State University, Bukirev
Street 15, 614990 Perm, Russia}
\affiliation{Department of Control Theory, Nizhny Novgorod State University, Gagarin Avenue 23, 603022 Nizhny Novgorod, Russia}
\email{denis.goldobin@gmail.com}
\date{\today}
\begin{abstract}
We report the effect of nonlinear bias of the frequency of collective oscillations of sin-coupled phase oscillators subject to individual asymmetric Cauchy noises. The noise asymmetry makes the Ott--Antonsen Ansatz inapplicable. We argue that, for all stable non-Gaussian noises, the tail asymmetry is not only possible (in addition to the trivial shift of the distribution median) but also generic in many physical and biophysical set-ups.
For the theoretical description of the effect, we develop a mathematical formalism based on the circular cumulants. The derivation of rigorous asymptotic results can be performed on this basis but seems infeasible in traditional terms of the circular moments (the Kuramoto--Daido order parameters). The effect of the entrainment of individual oscillator frequencies by the global oscillations is also reported in detail. The accuracy of theoretical results based on the low dimensional circular cumulant reductions is validated with the high-accuracy ``exact'' solutions calculated with the continued fraction method.
\end{abstract}

\maketitle

\begin{quotation}
The macroscopic dynamics of many paradigmatic models of nonlinear oscillator populations, including neural ones, are low dimensional and can be exactly described within the framework of the Ott--Antonsen Ansatz for the case of Cauchy noise. Not surprisingly, this mathematical framework was repeatedly employed for theoretical studies.
However, in contrast to the Gaussian noise, which cannot be asymmetric, all other stable noises, including the Cauchy one, can be asymmetric.
The asymmetry of microscopic fluctuations, which give rise to the effective white macroscopic noise, creates merely a shift of the mean value of Gaussian noise in the absence of heavy-tailed large fluctuations, but it results in an alpha-stable (L\'evy) noise with asymmetric tails in the presence of these large fluctuations.
The asymmetry of non-Gaussian stable noises was well established and appreciated in financial statistics by Mandelbrot and other researchers.
For physical and biophysical systems, in many generic set-ups, one ought to expect an asymmetric Cauchy noise, which violates the applicability conditions of the Ott--Antonsen Ansatz. In this paper we provide a mathematical framework for theoretical description of the macroscopic dynamics of oscillator populations subject to asymmetric Cauchy noise; the framework is based on the formalism of so-called ``circular cumulants.''
Further, we report the phenomenon of nonlinear bias of collective oscillation frequency, which is completely discarded by the Ott--Antonsen Ansatz, by the noise asymmetry for a Kuramoto-type ensemble. In networks of heterogeneous oscillators, the effect of frequency entrainment of individual oscillators by the global oscillation can be quite divers and is theoretically characterized within the developed framework.
\end{quotation}


\section{Introduction}

In the theory of collective phenomenon an important and broad class of paradigmatic models of phase oscillator populations was found to exhibit low dimensional macroscopic dynamics. The mathematical nature of these dynamics was elucidated by the Watanabe--Strogatz theory;\cite{Watanabe-Strogatz-1993,Watanabe-Strogatz-1994,Pikovsky-Rosenblum-2008,Marvel-Mirollo-Strogatz-2009} later on, the Ott--Antonsen (OA) theory delivered explicit low dimensional equations for the dynamics of the Kuramoto order parameter.\cite{Ott-Antonsen-2008,Ott-Antonsen-2009} This theory allowed for a significant theoretical advance with understanding many collective phenomenon.\cite{Abrams-etal-2008,Nagai-Kori-2010,Braun-etal-2012,Pazo-Montbrio-2014,Laing-2015,Pimenova-etal-2016,Dolmatova-Goldobin-Pikovsky-2017,Omelchenko-Knobloch-2019,Klinshov-Franovic-2019,Yasmine-etal-2022} In particular, it gave birth to the ``next generation neural mass models'' in mathematical neuroscience.\cite{Pazo-Montbrio-2014,Laing-2015,Coombes-2019,Coombes-2023}

For awhile, there was a persisting challenge of the generalization of the original OA theory to imperfect situations, where the applicability conditions of the OA theory are slightly violated. Arguably, the most important of all might be the case of individual noise. For instance, endogenous noise was identified to be responsible for the emergence of collective gamma-band oscillations in balanced networks of synaptically coupled neurons.\cite{Volo-Torcini-2018}. The effect was also observable for the quadratic integrate-and-fire model (QIF) of neurons, for which the OA theory is valid in the noise-free case. However, the OA model reduction (not a rigorous theory in this case) does not show appearance of macroscopic self-oscillations in the presence of noise.\cite{Volo-Torcini-2018}
The generalization of the OA theory on the basis of the so-called ``circular cumulants''\cite{Tyulkina-etal-2018,Goldobin-Dolmatova-2019b} allowed to construct low-dimensional
model reductions incorporating the effect of extrinsic and endogenous noises\cite{Ratas-Pyragas-2019,diVolo-etal-2022,Zheng-Kotani-Jimbo-2021,Goldobin-2021} and mathematical theoretical descriptions for macroscopic collective oscillations in networks of QIFs. These macroscopic theories employed the ``diffusion approximation'' for noise, which assumes the noise to be white and Gaussian.\cite{Capocelli-Ricciardi-1971,Tuckwell-1988}

The Lorentzian noise is an exceptional case, for which the OA theory is exact. Recently,\cite{Toenjes-Pikovsky-2020,Pietras-etal-2023,Pyragas2-2023,Kostin-etal-2023} the macroscopic dynamics of oscillator populations with Lorentzian (symmetric Cauchy) noise were extensively studied by means of the OA Ansatz.
In particular, the effect of Lorentzian noise for QIFs was found to be practically equivalent to the one of quenched disorder (frozen heterogeneity of natural frequency, coupling, etc.); no analogue of the aforementioned collective oscillations induced by endogenous Gaussian noise was reported.
However, dealing with Cauchy noise, one must be aware of an issue which is simply impossible for (and only for) Gaussian noise while is abundant for all non-Gaussian noises.  Here one ought to make one step back and discuss the physical origin of the mathematical models of white Gaussian and non-Gaussian noises.

Consider microscopic fluctuations with correlation time much shorter than the time scale of the macroscopic system dynamics. One can take a mesoscopic time interval long compared to the fluctuation correlation time and short compared to the macroscopic time scale. The impact of fluctuations on this time interval will be given by the sum of a large number of independent microscopic fluctuations and the impacts on successive mesoscopic intervals will be independent. Thus, we arrive to an effective noise, which is ``continuous'' and $\delta$-correlated on the mesoscopic time scale. Simultaneously, it is distributed as the sums of large number of independent random numbers. According to the Central limit theorem, if the distribution of summands (microscopic fluctuations) decays as $1/|x|^3$ or faster, these sums are Gaussian distributed.\cite{Zolotarev-1986} This is the ground for the stochastic mathematical models with $\delta$-correlated Gaussian noise. In particular, in the mathematical neuroscience, the aforementioned ``diffusion approximation'' is adopted not only for extrinsic noises but also for the endogenous noise generated by the irregularity of incoming synaptic pulse trains from other neurons of the network, given sufficiently large number of inbound synapses.\cite{diVolo-etal-2022,Goldobin-Volo-Torcini-2024}

However, if the distribution of microscopic fluctuations possesses a heavier tails $1/|x|^{\alpha+1}$ with $0<\alpha<2$, the sums will be distributed with the power-law tails of the same exponent (distributions with $\alpha\le0$ are impossible as they cannot be normalized). Such heavy-tailed fluctuations are also called L\'evy flights and not exotic in physics, geology, biology, and finances.\cite{Klafter-Blumen-Shlesinger-1987,Goldobin-Dolmatova-2019,Petrovskii-Morozov-2009,Mandelbrot-1963,Simkowitz-Beedles-1980} In particular, the Lorentzian distribution corresponds to $\alpha=1$ and this distribution is quite abundant for disorder in physical systems.\cite{Rabinovich-Trubetskov-1989,Lamb-1964,Lloyd-1969,Goldobin-Pikovsky-2005} Therefore, the case of white Cauchy noise is physically meaningful and receives a deserved attention from researchers.

In Sec.~\ref{sec:ASN}, with Eqs.~(\ref{eq101}) and (\ref{eq102}), we will discuss one of unique properties of Gaussian noise which distinguishes it from {\it all} heavy-tailed L\'evy noises: it cannot be asymmetric (skewed). Asymmetry of microscopic fluctuations forming the noise merely shifts its mean value.
In particular, for the endogenous noise generated by the irregularity of the arrival timing of inhibitory synaptic pulses of constant amplitude, the effective fluctuations of the synaptic current around the time-averaged value are Gaussian and possess a symmetric distribution.\cite{Capocelli-Ricciardi-1971,Tuckwell-1988,diVolo-etal-2022}
On the contrary, for heavy-tailed microscopic fluctuations, their inherent asymmetry creates an asymmetric L\'evy fluctuations of synaptic current. The asymmetric L\'evy noise turns out to be not merely possible but natural for many physically and biologically relevant set-ups.

Recently, triggered by Ref.~\citen{Roberts-etal-2015}, the impact of L\'evy noise on collective dynamics in neural and other systems received a revitalized attention from the nonlinear dynamics community.\cite{Wang-etal-2021,Wang-etal-2022,Wang-etal-2022,Dolmatova-Tyulkina-Goldobin-2023,Li-etal-2016,Rybalova-Nikishina-Strelkova-2024,Rybalova-etal-2024b,Rybalova-etal-2024c,Goldobin-2024}

In this paper, we report the effect of the frequency bias of collective oscillations induced by asymmetric Cauchy noise for the Kuramoto ensemble. This effect will be shown to be essentially nonlinear. It completely drops out within the framework of the Ott--Antonsen Ansatz, which is accurate for a symmetric Cauchy noise, but can be theoretically described with the formalism of circular cumulants.

The paper is organized as follows. In Sec.~\ref{sec:ASN}, we provide mathematical preliminaries for asymmetric L\'evy noises and rigorous motivation of the interest to them. In Sec.~\ref{ssec:CCR}, we derive the circular cumulant reduction equations for a population of phase oscillators driven by independent $\delta$-cor\-re\-lated asymmetric Cauchy noises. In Sec.~\ref{ssec:ti}, we construct a rigorous asymptotic expansion for the time-independent macroscopic states and discuss the convergence of circular cumulant expansions for such states. In Sec.~\ref{sec:entrain}, the problem of individual frequency entrainment for heterogeneous populations with global coupling is solved within the low dimensional circular cumulant reductions.
In Sec.~\ref{sec:bias}, the effect of nonlinear bias of collective oscillation frequency by the noise asymmetry is presented and theoretically explained for the Kuramoto ensemble. In Sec.~\ref{ssec:entrainKur}, the entrainment of individual frequencies is characterized. In Sec.~\ref{sec:contfrac}, the employment of the continued fraction method for high-precision numerical solutions for the Kuramoto ensemble is described.
Conclusions are summarized in Sec.~\ref{sec:concl}.

\section{Mathematical preliminaries: alpha-stable distribution skewness}\label{sec:ASN}
As the number of elements of the sum of independent identically-distributed random numbers grows, the distribution of the sum tends to an invariant limiting shape, which is called ``$\alpha$-stable distribution.''\cite{Zolotarev-1986} Gaussian and Lorentzian distributions are the only cases with a simple analytical formula of the distribution. The stable distributions $w(x)$ are naturally given by their characteristic functions
\begin{equation}
F_x(k)=\langle{e^{ikx}}\rangle=\exp\big[ik\mu-|ck|^\alpha(1+i\beta\mathrm{sign}(k)\Theta)\big]\,,
\label{eq101}
\end{equation}
where $\alpha\in(0,2]$, $\mu\in\mathbb{R}$ is a shift parameter, $c>0$ is a scale parameter measuring the distribution width, $\beta\in[-1,1]$ is the skewness parameter,
\begin{equation}
\Theta=\left\{
\begin{array}{cc}
\tan\left(\frac{\pi\alpha}{2}\right), & \mbox{ for } \alpha\ne1; \\
-\frac{2}{\pi}\ln|k|, & \mbox{ for } \alpha=1.
\end{array}
\right.
\label{eq102}
\end{equation}
For $\alpha=2$, Eq.~(\ref{eq101}) yields the characteristic function of a Gaussian distribution; for $\alpha=1$ and $\beta=0$, this is a characteristic function of the Lorentzian (symmetric Cauchy) distribution with median $\mu$ and half-width at half-height $c$.
The sum of two independent $\alpha$-stable random variables is an $\alpha$-stable random variable with parameters 
\begin{align}
\mu&=\mu_1+\mu_2\,,
\label{eq103}
\\
|c|&=(|c_1|^\alpha+|c_2|^\alpha)^{1/\alpha}\,,
\label{eq104}
\\
\beta&=\frac{\beta_1|c_1|^\alpha+\beta_2|c_2|^\alpha}{|c_1|^\alpha+|c_2|^\alpha}\,.
\label{eq105}
\end{align}

For a correct generation of the discrete time version of a $\delta$-correlated $\alpha$-stable noise, one has to take the following properties into account.
For the sum of two noise increments $\xi_{\Delta{t}}\Delta{t}$ the equivalent increment for the time step size $(2\Delta{t})$ is $\xi_{2\Delta{t}}2\Delta{t}=2^{1/\alpha}\xi_{\Delta{t}}\Delta{t}$ [see Eq.~(\ref{eq104})]; therefore, $\xi_{\Delta{t}}\propto(\Delta{t})^{1/\alpha-1}$. One must comply with the latter scaling law in numerical simulations and in the derivations of continuous-time mathematical description from more intuitively obvious discrete time considerations.\cite{Dolmatova-Tyulkina-Goldobin-2023}

With Eq.~(\ref{eq102}), one can see the unique property of the case of $\alpha=2$: a Gaussian distribution cannot be asymmetric, the asymmetry of summands merely shifts $\mu$. For $\alpha\ne2$, the summands with asymmetric tails generate an asymmetric $\alpha$-stable distribution.

\begin{figure}[!t]
\center{
\includegraphics[width=0.475\textwidth]%
 {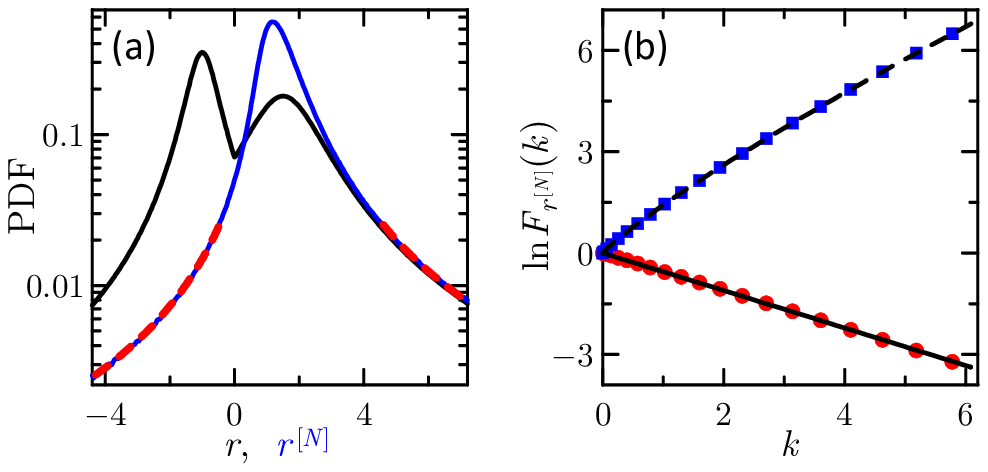}
}
\caption{Formation of an asymmetric $\alpha$-stable distribution is illustrated with large sums for $\alpha=1$.
(a):~Black line: a piecewise Lorentzian distribution of independent random numbers $r_j$ with a continuous PDF $C_1/[(r-x_1)^2+\Delta_1^2]$ for $r<0$ and $C_2/[(r-x_2)^2+\Delta_2^2]$ for $r\ge0$, where $(x_1,\Delta_1,x_2,\Delta_2)=(-1,0.5,1.5,1.2)$ and $C_{1,2}$ are the normalization constants;
blue line: the PDF of sums $r^{[N]}=N^{-1}\sum_{j=1}^Nr_j$ for large $N=1000$;
red dashed lines: the asymptotic laws for the tails of the PDF of sums, $a_n/(r^{[N]}-\mu_N)^2$, where $a_1=0.081$ ($a_2=0.271$) for the left (right) tail and $\mu_N$ is taken from panel~(b).
(b):~The real (imaginary) part of the logarithm of the numerically calculated characteristic function of $r^{[N]}$ is plotted with red circles (blue squares); the solid (dashed) line shows the fitting of  $\ln{F_{r^{[N]}}(k)}$ with $i\mu_Nk-\sigma_N(|k|-i[2\beta_N/\pi]k\ln|k|)$, where
$\mu_N=1.422$, $\sigma_N=0.555$, $\beta_N=0.483$.
}
  \label{fig1}
\end{figure}

\begin{figure}[!t]
\center{
\includegraphics[width=0.475\textwidth]%
 {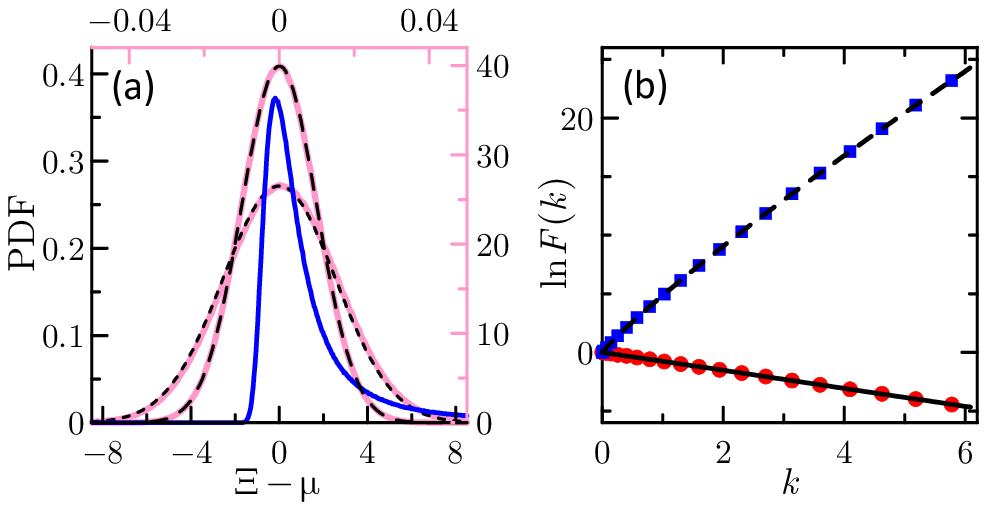}
}
\caption{The nature of the ``diffusion approximation'' for Poissonian processes of $\delta$-pulses $\xi(t)$ is illustrated with the PDF of the integral $\Xi=\int_0^1\xi(t)\mathrm{d}t$ for the rate of pulses arrival $10^4$. (a):~Pink lines: the PDFs for pulses of strength $1$ and pulses of random strength $\zeta_j$ with PDF $P(\zeta)=C_\zeta\zeta^2/[(\zeta^3-x_2)^2+\Delta_2^2]$ for $\zeta\ge0$ and $0$ for $\zeta<0$, the dashed and short-dashed black lines show the Gaussian fitting for the unit and $\zeta_j$ pulses, respectively, see the right and upper pink axes for the scale;
blue line: the PDF for pulses with heavy tail of $P(\zeta)=B_\zeta/[(\zeta-x_2)^2+\Delta_2^2]$ for $\zeta\ge0$ and $0$ for $\zeta<0$. Parameters: $x_2=1.5$, $\Delta_2=1.2$. In the latter case, the distribution of $\Xi$ is non-Gaussian and asymmetric: one cannot use an effective continuous Gaussian noise approximation for the limit of large pulse arrival rate. (b):~For the latter case, the logarithm of the characteristic function of $\Xi$ is fitted with $\mu_N=4.87$, $\sigma_N=0.77$, $\beta_N=1$ (notations are the same as in Fig.~\ref{fig1}b).
}
  \label{fig2}
\end{figure}

In Fig.~\ref{fig1}, we illustrate that the sums of asymmetric microscopic fluctuations with Cauchy tails generate an asymmetric Cauchy-type distribution and this distribution is $\alpha$-stable, which is validated with the characteristic function in Fig.~\ref{fig1}b perfectly fitted with shape (\ref{eq101}). Further, we take a shot noise, which, for instance, one would find for trains of incoming synaptic pulses in neuronal populations with random networks of synaptic links\cite{diVolo-etal-2022} and finite globally coupled subpopulations:\cite{Klinshov-Kirillov-2022,Klinshov-Smelov-Kirillov-2023}
\[
s(t)=\sum_j\zeta_{j}\delta(t-t_j)\,,
\]
where $t_j$ and $\zeta_j$ are the arrival time and strength of the $j$th pulse, respectively. In the case of a sparse network of synaptic links, the arrival times are uncorrelated on the mesoscopic time scale and one can think of a Poissonian process.\cite{diVolo-etal-2022,Goldobin-Volo-Torcini-2024} If the pulses are of identical strength the integral over a mesoscopic time interval
\[
\Xi_\tau=\int_0^\tau s(t)\,\mathrm{d}t
\]
will be Gaussian-distributed. Therefore, given the number of incoming pulses per $\tau$ is large, one can effectively represent the input drive as $s(t)=\langle{s(t)}\rangle+\sigma_s\xi(t)$, where $\xi(t)$ is a white Gaussian noise (see Fig.~\ref{fig2}a). This gives rise to the so-called ``diffusion approximation'' broadly accepted and well recommended itself in mathematical neuroscience.\cite{Capocelli-Ricciardi-1971,Tuckwell-1988}

If one considers distributed $\zeta_j$, which corresponds to a nonidentical strength of synaptic links, and their distribution possesses tails decaying non-slower than $\propto1/|\zeta|^3$, the effective noise will be still Gaussian. In Fig.~\ref{fig2}a, one can witness a Gaussian distribution for the case of $1/|\zeta|^4$-tails; the convergence to the Gaussian distribution is slow in this case and one can notice a remaining tiny asymmetry of the distribution (the asymmetry vanishes for longer sums). However, for heavier tails of the distribution of $\zeta_j$, we arrive to the case of an asymmetric $\alpha$-stable noise.
The researchers thoroughly studied the case of a Lorentzian quenched disorder or/and noise.\cite{Pazo-Montbrio-2014,Pietras-etal-2023,Pyragas2-2023} Given we take a Lorentzian distribution of $\zeta_j$ and truncate all negative ({\it or} positive) links due to a biological reason that one should expect different mechanisms of formation for excitory and inhibitory links and, therefore, nonidentical distributions of the strength of two, we arrive to the case of an asymmetric Cauchy distribution (see Fig.~\ref{fig2}a and b).
The argument of the different distributions for positive and negative links makes also the first illustration (Fig.~\ref{fig1}) biophysically relevant.

\section{Macroscopic equations for populations with asymmetric Cauchy noise}\label{sec:CCs}
\subsection{Circular cumulant reduction for the population dynamics under asymmetric Cauchy noise}\label{ssec:CCR}
It is instructive to start from finite circular cumulant reductions and derive the infinite chain of CC equations in Sec.~\ref{ssec:CCChain}, since this generalized derivation is much more involved. The results of entire Sec.~\ref{sec:CCs} are not restricted to the case of the Kuramoto ensemble and are applicable to a wide class of ensembles.
Namely, we consider a population of sin-coupled phase elements (oscillators) subject to additive $\delta$-correlated noises $\sigma\xi_j(t)$:
\begin{equation}
\dot\varphi_j=\omega(t)+\mathrm{Im}(2h(t)e^{-i\varphi_j})+\sigma\xi_j(t)\,,
\label{eq110a}
\end{equation}
where $\xi_j(t)$ are drawn from an asymmetric Cauchy distribution ($\alpha=1$, $\beta\ne0$).
This shape of the $h$-term covers many paradigmatic mean-field coupling models from the theory of collective phenomena; in particular, for the Kuramoto coupling of strength $\varepsilon$ [see Eq.~(\ref{eq201})] $h=\varepsilon Z_1/2$, where the Kuramoto order parameter $Z_1=N^{-1}\sum_{j=1}^N\exp[i\varphi_j]$, for a shunted chain of superconducting Josephson junctions $h$ is a constant proportional to the shunting Ohmic resistance,\cite{Watanabe-Strogatz-1994} the synaptic coupling of theta-neurons can be also conveyed by such form.~\cite{Pazo-Montbrio-2014,Laing-2015}

One can derive a fractional Fokker--Planck equation governing the dynamics of the average probability density $w(\varphi,t)$ (e.g., see Refs.~\citen{Chechkin-etal-2003,Toenjes-etal-2013} or \citen{Goldobin-Permyakova-Klimenko-2024} for the formulations in terms of this paper). In the Fourier space, where
\begin{equation}
w(\varphi,t)=\frac{1}{2\pi}\sum_{m=-\infty}^{+\infty}a_m(t)e^{-im\varphi},
\label{eq:wFour}
\end{equation}
$a_0=1$ and $a_{-m}=a_m^\ast$, the Fokker--Planck equation reads
\begin{equation}
\dot{a}_m=im\omega a_m+mh\,a_{m-1}-mh^\ast a_{m+1} +\sigma\dot\Phi^{(\xi)}(m)\, a_m\,,
\label{eq113}
\end{equation}
where
\[
\dot\Phi^{(\xi)}(m)\equiv\frac{\Phi_{\xi\Delta t}\!\!\left(m;c=\Delta t\right)}{\Delta t}
\]
and $\Phi_{\xi\Delta t}(k;c)\equiv\ln F_{\xi\Delta t}(k;c)$ is the logarithm of the characteristic function (\ref{eq101}) for the discrete time version of noise $\xi(t)$ and time step $\Delta t$ (see Appendix~\ref{app:FFP} for derivation).
For an asymmetric Cauchy noise [see Eqs.~(\ref{eq101}) and (\ref{eq102})],
\begin{equation}
\dot\Phi^{(\xi)}(m)=-|m|+i\frac{2\beta}{\pi}m\ln|m|\,.
\label{eq113Phi}
\end{equation}




One can also include here the case of a heterogeneous population with a Lorentzian distribution of natural frequencies
\[
G(\omega)=\frac{\pi^{-1}\gamma}{\gamma^2+(\omega-\omega_0)^2}\,,
\]
which yields equations for the Kuramoto--Daido order parameters
$Z_m(t)=\int G(\omega)\,a_m(\omega,t)\,\mathrm{d}\omega$
from Eq.~(\ref{eq113}):
\begin{equation}
\dot{Z}_m=m\big((i\omega_0-\gamma)Z_m+hZ_{m-1}-h^\ast Z_{m+1}\big) +\sigma\dot\Phi^{(\xi)}(m)\,Z_m\,,
\label{eq114}
\end{equation}
where $Z_0=1$.

One can introduce ``circular cumulants''~\cite{Tyulkina-etal-2018,Goldobin-etal-2018,Goldobin-2019,Goldobin-Dolmatova-2019b} (CCs) related to $Z_m$ via the generating function. The moment-generating function is an analogue of the conventional characteristic function, but for a cyclic variable:
\begin{equation}
F(k,t)=\langle e^{ke^{i\varphi}}\rangle=\sum_{m=0}^{\infty}Z_m(t)\frac{k^m}{m!}\,,
\label{eqFkt}
\end{equation}
and its logarithm provides the CC-generating function
\begin{equation}
\Psi(k,t)=k\frac{\partial}{\partial k}\ln{F(k,t)}=\sum_{m=1}^{\infty}\kappa_mk^m.
\label{eqPSIkt}
\end{equation}
The first three CCs are
\begin{align}
\kappa_1=Z_1\,,
\qquad
\kappa_2=Z_2-Z_1^2,
\qquad
\kappa_3=\frac{Z_3-3Z_2Z_1+2Z_1^3}{2}\,;
\nonumber
\end{align}
for arbitrary order, one can use recursive formulae:
$$
\kappa_m=\frac{Z_m}{(m-1)!}-\sum_{n=1}^{m-1}\frac{\kappa_nZ_{m-n}}{(m-n)!}
\,.
$$
For the thirst three CCs one can write $\dot\kappa_1=\dot{Z}_1$, $\dot\kappa_2=\dot{Z}_2-2Z_1\dot{Z}_1$, etc., and derive the chain of the dynamics equations for circular cumulants:
\begin{align}
\dot{Z}_1&=(i\omega_0-\gamma-\sigma)Z_1 +h-h^\ast(\kappa_2+Z_1^2)\,,
\label{eq115}
\\[5pt]
\dot{\kappa}_2&=2(i\omega_0-\gamma-\sigma)\kappa_2 -4h^\ast(\kappa_3+Z_1\kappa_2)
\nonumber\\
&\qquad\qquad\qquad\qquad
+i\frac{2\beta\sigma}{\pi}\ln{4}(\kappa_2+Z_1^2)\,,
\label{eq116}
\\[5pt]
\dot{\kappa}_3&=3(i\omega_0-\gamma-\sigma)\kappa_3 -h^\ast(9\kappa_4+6Z_1\kappa_3+3\kappa_2^2) \nonumber\\
&\qquad
+i\frac{3\beta\sigma}{\pi}\left(\ln{9}\kappa_3
+\ln\frac{27}{4}\kappa_2Z_1 +\ln\frac{3}{4}Z_1^3
\right)\,.
\label{eq117}
\end{align}
In Sec.~\ref{ssec:CCChain}, the infinite chain of the CC equations is derived, but Eqs.~(\ref{eq115})--(\ref{eq117}) are already sufficient for us to explain some general observations, which can be also seen to be valid with the infinite chain given by~(\ref{eqap02}) and (\ref{eqap06}).


For $\beta=0$ (symmetric Cauchy noise) or $\sigma=0$ (noise-free), the equation chain~(\ref{eq115})--(\ref{eq117}) admits solution $Z_1\ne0$, $\kappa_{m\ge2}=0$, which is the OA Ansatz, where the dynamics of $Z_1$ is governed solely by Eq.~(\ref{eq115}); moreover, this solution is attracting~\cite{Ott-Antonsen-2009} for $\gamma+\sigma>0$. Hence, the CC representation has potential to be a useful framework for studying nonideal situations, where higher $\kappa_m$ are nonzero but small.

For $\beta\sigma>0$, Eqs.~(\ref{eq116}) and (\ref{eq117}) yield
\begin{align}
\kappa_2&\propto\beta\sigma Z_1^2\ln{4}\,,
\nonumber\\
\kappa_3&\propto\beta\sigma\left(\ln\frac{3}{4}Z_1^3
+\ln\frac{27}{4}\kappa_2Z_1\right) +\mathcal{O}\big(\kappa_2^2\big)\,.
\nonumber
\end{align}
Here two different smallness hierarchies can emerge:
\\
(i)~for a weak collective mode $|Z_1|\ll1$, circular cumulants $\kappa_2\propto Z_1^2$, $\kappa_3\propto Z_1^3$, \dots $\kappa_m\propto Z_1^m$;
\\
(ii)~for weak noise $\sigma\ll1$ or small noise asymmetry $\beta\ll1$, circular cumulants $\kappa_2\propto(\beta\sigma\ln{4})Z_1^2$ and $\kappa_3\propto\beta\sigma(\ln\frac{3}{4}Z_1^3+\ln\frac{27}{4}\kappa_2Z_1)$, whence one can estimate the order of magnitude of the ratio
$\kappa_3/\kappa_2\sim[\ln(3/4)/\ln{4}]Z_1+[\ln(27/4)/\ln{4}]\kappa_2/Z_1+\mathcal{O}(\beta\sigma)\approx-0.2Z_1$.
\\
In case~(i), we have an obvious decaying geometric progression of $\kappa_m$. In case~(ii), the second CC makes a correction to the OA solution, but $\kappa_3$ is of the same order of smallness. However, first, $\kappa_3$ makes an indirect impact on the dynamics of $Z_1$, only via the dynamics of $\kappa_2$, which is immediately present in Eq.~(\ref{eq115}) for $Z_1$; second, $|\kappa_3/\kappa_2|$ is smaller than $(1/5)|Z_1|$, while $|Z_1|\le1$. Hence, $\kappa_3$ can be expected to be often of minor significance against the background of the $\kappa_2$-correction. In summary, one can adopt an approximate closure $\kappa_3=0$ as a result of rigorous asymptotic expansion in some cases or as a rough approximation in other circumstances.


The geometric interpretation of the two first CCs for a phase variable was explained via analogy with a variable on the infinite line.~\cite{Goldobin-Dolmatova-2019b}
For a phase variable, the first circular cumulant (the Kuramoto order parameter) characterizes the location of the distribution center with $\arg{Z_1}$ and its width with $\ln(1/|Z_1|)$.
The second circular cumulant quantifies the distribution asymmetry with $(\arg\kappa_2-2\arg{Z_1})$ and the deformation of tails with $|\kappa_2|$, which are analogues of skewness and kurtosis for a variable on the line, respectively.
The third circular cumulant provides characterization beyond the analogues of skewness and kurtosis.

\subsection{Circular cumulant equation chain}\label{ssec:CCChain}
The infinite chain of CC equations can be derived via the dynamics of the moment-generating function~(\ref{eqFkt}) of phase $\varphi$. One  calculates~\cite{Tyulkina-etal-2018,Zheng-Kotani-Jimbo-2021,Goldobin-2021} 
the time-derivative
\begin{align}
&\frac{\partial}{\partial t}F(k,t)=\sum_{m=0}^{\infty}\dot{Z}_m(t)\frac{k^m}{m!}
\nonumber\\
&=(i\omega_0-\gamma)k\frac{\partial}{\partial k} F+hkF-h^\ast k\frac{\partial^2}{\partial k^2}F +\sigma\dot\Phi^{(\xi)}\left(k\frac{\partial}{\partial k}\right)F\,,
\nonumber
\end{align}
where $\dot{Z}_m$ are substituted from Eq.~(\ref{eq114}). As
$\Psi(k,t)=\frac{k}{F}\frac{\partial F}{\partial k}
=k\frac{\partial}{\partial k}\ln{F(k,t)}$ (\ref{eqPSIkt}), one finds $\dot{\Psi}=k\frac{\partial}{\partial k}\frac{\dot{F}}{F}$ and
\begin{align}
\frac{\partial\Psi}{\partial t}
&=(i\omega_0-\gamma-\sigma)k\frac{\partial}{\partial k}\Psi+hk
 -h^\ast k\frac{\partial}{\partial k}
 \left(k\frac{\partial}{\partial k}\frac{\Psi}{k}+\frac{\Psi^2}{k}\right)
\nonumber\\
&\qquad\qquad
{}+i\frac{2\beta\sigma}{\pi} k\frac{\partial}{\partial k}\left(
 \frac{1}{F}k\frac{\partial}{\partial k}\ln\left(k\frac{\partial}{\partial k}\right) F\right)\,.
\label{eqap01}
\end{align}
Substituting series~(\ref{eqPSIkt}) into (\ref{eqap01}) and collecting terms with $k^m$, one finds
\begin{align}
\dot\kappa_m&=m(i\omega_0-\gamma-\sigma)\kappa_m+h\delta_{m\,1}
\nonumber\\
&{}
 -h^\ast\Big(m^2\kappa_{m+1}+m\sum_{n=1}^{m}\kappa_{m-n+1}\kappa_{n}\Big)
 +im\frac{2\beta\sigma}{\pi}\mathcal{G}_m^{(\beta)}\,,
\label{eqap02}
\end{align}
where $\mathcal{G}_m^{(\beta)}$ are the coefficients of the power series of
\[
\mathcal{G}^{(\beta)}\equiv\frac{1}{F}k\frac{\partial}{\partial k}\ln\left(k\frac{\partial}{\partial k}\right)F
\equiv\sum_{m=1}^{\infty}\mathcal{G}_m^{(\beta)}k^m.
\]
The coefficients of the series
\[
\mathcal{G}^{(\beta)}=e^{-\sum_{n=1}^{\infty}\kappa_n\frac{k^n}{n}}
k\frac{\partial}{\partial k}\ln\left(k\frac{\partial}{\partial k}\right)
e^{\sum_{l=1}^{\infty}\kappa_l\frac{k^l}{l}},
\]
where $k\frac{\partial}{\partial k}\ln(k\frac{\partial}{\partial k})$ does not change the exponent of $k^m$ but creates the multiplier $m\ln{m}$ for the $k^m$-term of the right-hand-side exponential, does not seem to be representable by an explicit formula.

Using the recursive forward and backward formulae for circular cumulants and moments
\begin{align}
\kappa_m&=\frac{Z_m}{(m-1)!}-\sum_{n=1}^{m-1}\frac{\kappa_nZ_{m-n}}{(m-n)!}\,,
\label{eqap03}
\\
Z_m&=(m-1)!\kappa_m+\sum_{n=1}^{m-1}\frac{(m-1)!}{(m-n)!}\kappa_nZ_{m-n}\,,
\label{eqap04}
\end{align}
one can write the time-derivative of (\ref{eqap03}),
$
\dot\kappa_m=\frac{\dot{Z}_m}{(m-1)!}-\sum_{n=1}^{m-1}\frac{\kappa_n\dot{Z}_{m-n}+\dot\kappa_nZ_{m-n}}{(m-n)!}
$,
and obtain a recursive formula
\begin{equation}
m\mathcal{G}_m^{(\beta)}=\frac{m\ln{m} Z_m}{(m-1)!}
-\sum_{n=1}^{m-1}\frac{\big(\kappa_n(m-n)\ln(m-n)+n\mathcal{G}_n^{(\beta)}\big)Z_{m-n}}{(m-n)!}\,,
\label{eqap05}
\end{equation}
where $n\ln{n}=0$ for $n=0$ and $1$.
Starting from $m=1$, one can recursively employ Eqs.~(\ref{eqap04}) and (\ref{eqap05}) to calculate $\mathcal{G}_m^{(\beta)}$ as a function of $\kappa_n$ with $n=1,...,m$ for all $m$.

\begin{figure}[!t]
\center{
\includegraphics[width=0.275\textwidth]%
 {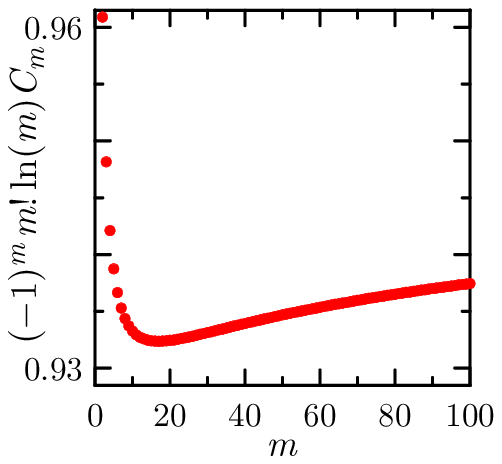}
}
\caption{The ratio of $C_m$ to approximation (\ref{eq:appCm}) is plotted {\it vs} $m$.
}
  \label{fig3}
\end{figure}

Employing the recursive procedure, one obtains
\begin{subequations}
\label{eqap06}
\begin{align}
\mathcal{G}_1^{(\beta)}&=0\,,
\label{eqap06a}
\\
2\mathcal{G}_2^{(\beta)}&=2\ln{2}(\kappa_2+\kappa_1^2)\,,
\label{eqap06b}
\\
2\mathcal{G}_3^{(\beta)}&=2\ln{3}\kappa_3+\ln\frac{27}{4}\kappa_2\kappa_1 +\ln\frac{3}{4}\kappa_1^3\,,
\label{eqap06c}
\\
2\mathcal{G}_4^{(\beta)}&=2\ln{4}\kappa_4 +\frac{2}{3}\ln\frac{256}{27}\kappa_3\kappa_1 +\ln{2}\kappa_2^2
\nonumber\\
&
{}+\ln\frac{16}{27}\kappa_2\kappa_1^2 +\frac{1}{3}\ln\frac{32}{27}\kappa_1^4\,,
\label{eqap06d}
\\
2\mathcal{G}_5^{(\beta)}&=2\ln{5}\kappa_5+\frac{1}{2}\ln\frac{5^5}{4^4}\kappa_4\kappa_1 +\frac{1}{3}\ln\frac{5^5}{2^2\times3^3}\kappa_3\kappa_2
\nonumber\\
&
{}+\frac{1}{3}\ln\frac{3^3\times5^5}{8^6}\kappa_3\kappa_1^2
+\frac{1}{4}\ln\frac{5^5}{2^4\times3^6}\kappa_2^2\kappa_1
\nonumber\\
&
{}+\frac{1}{6}\ln\frac{5^5\times3^6}{4^{10}}\kappa_2\kappa_1^3
+\frac{1}{12}\ln\frac{5\times3^6}{4^6}\kappa_1^5\,,
\label{eqap06e}
\\
&\dots\;.
\nonumber
\end{align}
\end{subequations}
One can also derive the coefficient of the term $\kappa_1^m$ in $\mathcal{G}_m^{(\beta)}$, which will be required below for a rigorous asymptotic theory:
\begin{align}
\mathcal{G}_m^{(\beta)}&=C_m\kappa_1^m+\dots\,,
\nonumber
\\
C_m&=\sum_{n=0}^{m-2}\frac{(-1)^{n}\ln(m-n)}{n!(m-n-1)!}\,;
\label{eqap07b}
\end{align}
for instance, $C_1=0$, $C_2=\ln2$, $C_3=\frac12\ln3-\ln{2}=\frac12\ln\frac34$, $C_4=\frac{1}{3!}\ln{4}-\frac12\ln{3}+\frac{1}{2}\ln{2}=\frac{1}{6}\ln\frac{32}{27}$. An approximation
\begin{equation}
C_{m\ge2}\approx\frac{(-1)^m}{m!\ln{m}}
\label{eq:appCm}
\end{equation}
will be useful for the analysis of asymptotic expansions; the relative error of this approximation is below $7\%$ as one can see in Fig.~\ref{fig3}.

\subsection{Rigorous asymptotic dynamics of time-independent states}
\label{ssec:ti}
In this section we derive the rigorous perturbative equations for small $\beta\sigma$ and discuss their implications for the two-CCs approximation.

If $\gamma+\sigma>0$, the OA manifold ($\kappa_{m\ge2}=0$) is known to be attracting for $\beta\sigma=0$. Small but finite $\beta\sigma$ enforces deviations from the OA manifold; specifically, Eq.~(\ref{eqap02}) with $m\ge2$ and (\ref{eqap06}) dictates $\kappa_{m\ge2}\sim\beta\sigma Z_1^m$ [cf Eqs.~(\ref{eq115})--(\ref{eq117})]. For a finite $Z_1$, one should not neglect $\kappa_{m+1}$ in the dynamical equation for $\kappa_m$ as it is of the same order of magnitude with respect to $\beta\sigma$ as $\kappa_m$. To the first order in $\beta\sigma$, equation system~(\ref{eqap02}) simplifies to
\begin{align}
\dot{Z}_1&=(i\omega_0-\gamma-\sigma)Z_1 +h-h^\ast(\kappa_2+Z_1^2)\,,
\label{eq115_1}
\\[5pt]
\dot{\kappa}_m&=m(i\omega_0-\gamma-\sigma-2h^\ast Z_1)\kappa_m-m^2h^\ast\kappa_{m+1}
\nonumber\\
&\qquad
+im\frac{2\beta\sigma}{\pi}C_mZ_1^m\,,\quad
\mbox{ for } m=2,3,4,\dots\,.
\label{eq116_1}
\end{align}

Let us consider a time-independent state. Often, phase systems are invariant to the shift of all phases by the same value (e.g., Kuramoto and Kuramoto--Sakaguchi ensembles~\cite{Kuramoto-1975,Kuramoto-1984,Acebron-etal-2005,Sakaguchi-Kuramoto-1986}); in this case one can also have a ``rotating'' solution which is time-independent in a corotating frame. To include such situations in our study, we admit $\dot{\kappa}_m=im\Omega\kappa_m$, where the constant frequency $\Omega$ will be determined by the self-consistency conditions (is zero for the systems without the phase-shift invariance). With notation $\widetilde\omega=\Omega-\omega_0$, one can recast Eqs.~(\ref{eq115_1}) and (\ref{eq116_1}):
\begin{align}
&(\gamma+\sigma+i\widetilde\omega+h^\ast Z_1)Z_1-h=-h^\ast\kappa_2\,,
\label{eq115_ti}
\\[5pt]
&(\gamma+\sigma+i\widetilde\omega+2h^\ast Z_1)\kappa_m =-mh^\ast\kappa_{m+1}
+i\frac{2\beta\sigma}{\pi}C_mZ_1^m\,,
\label{eq116_ti}
\\
&\qquad\qquad
\mbox{ for }\quad
m=2,3,4,\dots\,.
\nonumber
\end{align}
The latter equation can be viewed as an explicit recursive relation $\kappa_m=ic_mZ_1^m-mb\kappa_{m+1}$, where
\[
c_m=\frac{2\beta\sigma\pi^{-1}C_m}{\gamma+\sigma+2h^\ast Z_1+i\widetilde\omega}\,,
\quad
b=\frac{h^\ast}{\gamma+\sigma+2h^\ast Z_1+i\widetilde\omega}\,.
\]
Hence, one can iteratively calculate
\begin{equation}
\kappa_m=\sum_{l=0}^\infty ic_{m+l}Z_1^{m+l}(-b)^l\frac{(m+l-1)!}{(m-1)!}
\end{equation}
for $m=2,3,\dots$\,. Specifically, for $\kappa_2$, which appears in the r.h.s.-part of Eq.~(\ref{eq115_ti}) for $Z_1$,
\begin{align}
\kappa_2&=i\frac{2\beta\sigma}{\pi}\frac{Z_1}{h^\ast}
\sum_{l=1}^\infty C_{l+1}(-1)^{l+1}\left(bZ_1\right)^ll!
\label{eq:kappa2ti}
\\
&\approx i\frac{2\beta\sigma}{\pi}\frac{Z_1}{h^\ast}
\sum_{l=1}^\infty \frac{1}{(l+1)\ln(l+1)}
\left(\frac{1}{\frac{\gamma+\sigma +i\widetilde\omega}{h^\ast Z_1}+2}\right)^l ,
\label{eq:kappa2ti_app}
\end{align}
where we employed approximate expression~(\ref{eq:appCm}) for the analysis of convergence of the series in (\ref{eq:kappa2ti}). Substituting $\kappa_2$ (\ref{eq:kappa2ti}) into Eq.~(\ref{eq115_ti}), one obtains a self-consistency equation for $Z_1$:
\begin{equation}
\gamma+\sigma+i\widetilde\omega+h^\ast Z_1-\frac{h}{Z_1}=
i\frac{2\beta\sigma}{\pi}
\sum_{l=1}^\infty(-1)^l l!\, C_{l+1}\left( bZ_1
\right)^l.
\label{eq:selfconsZ1}
\end{equation}

The convergence of this series is important as the $l$-th term of it is owned by $\kappa_{l+1}$; the truncation of the CC expansion after $\kappa_{l}$ is equivalent to the truncation of the series (\ref{eq:kappa2ti}) after $(l-1)$.
The magnitude of the geometric progression multiplier $(bZ_1)$ in brackets of (\ref{eq:kappa2ti}) can be further estimated to the leading order of (\ref{eq115_ti}) with respect to $\beta\sigma\ll1$: $\gamma+\sigma +i\widetilde\omega=h/Z_1-h^\ast Z_1+\mathcal{O}(\beta\sigma)$ and the multiplier becomes
\[
bZ_1\approx\frac{1}{1+\frac{h Z_1^\ast}{h^\ast Z_1}\frac{1}{|Z_1|^2}}\,.
\]
Notice, the absolute value of the second summand in the denominator is $\ge1$ and attains $1$ only for the perfect order states $|Z_1|=1$. For low synchrony states (small $|Z_1|$) the denominator is large and the series decays fast.

Let us consider the behavior of this multiplier in detail for a typical case --- Kuramoto--Sakaguchi global coupling~\cite{Sakaguchi-Kuramoto-1986}
\[
h=\frac{\varepsilon}{2}e^{-i\theta}Z_1\,,
\]
where $\varepsilon$ is the coupling strength and $\theta$ is the coupling phase shift ($\theta=0$: purely dissipative coupling, $\theta=\pm\pi/2$: purely conservative coupling, $\cos\theta<0$: repelling coupling).
For $\beta\sigma\to0$ or, as explained at the end of Sec.~\ref{ssec:CCR}, next to the synchronization  threshold, where $|\kappa_2|\propto |Z_1|^2\ll|Z_1|$, one can neglect the $\kappa_2$-term in Eq.~(\ref{eq115_ti}). In this case, the real part of Eq.~(\ref{eq115_ti}) divided by $Z_1$ becomes $\gamma+\sigma=\frac{\varepsilon}{2}\cos\theta(1-|Z_1|^2)$. The collective mode exists above the Kuramoto transition threshold, $\varepsilon>\varepsilon_\mathrm{cr}=2(\gamma+\sigma)/\cos\theta$, which is not affected by $\beta$, and
\[
\left.|Z_1|^2\right|_{\beta\sigma\to0}=1-\frac{\varepsilon_\mathrm{cr}}{\varepsilon}\,.
\]
In this case the multiplier $|bZ_1|\approx (\varepsilon-\varepsilon_\mathrm{cr})/[(\varepsilon-\varepsilon_\mathrm{cr})^2+\varepsilon^2 +2\varepsilon(\varepsilon-\varepsilon_\mathrm{cr})\cos{2\theta}]^{1/2}$ is smaller than $1$ for $\cos{2\theta}>-\frac{\varepsilon}{2(\varepsilon-\varepsilon_\mathrm{cr})}$. The latter condition is always fulfilled for $\varepsilon<2\varepsilon_\mathrm{cr}$; for higher coupling strength the condition is violated in a gap around $\theta=\pi/2$, i.e.\ for a mostly-conservative coupling with a small dissipative part. The latter case can be important: it might be related to the first analytically solvable example of ``Chimeras'' (solved by means of the OA Ansatz~\cite{Abrams-etal-2008}), where the coupling was slightly non-conservative (of course, an additional level of network connections in Ref.~\citen{Abrams-etal-2008} also complicates the situation significantly).

Generally, one can conclude that the case of a fast decay of the series in Eq.~(\ref{eq:kappa2ti}) is expected to be typical but not guarantied. The factor $(l+1)\ln(l+1)$ in the denominator of~(\ref{eq:kappa2ti_app}) assists the rate of convergence for $|bZ_1|\le1$ but does not prevent divergence for $|bZ_1|>1$.

\subsection{Entrainment of frequencies of individual oscillators}\label{sec:entrain}
Oscillations of an individual noisy oscillator can be more or less strongly entrained by the dynamics of the synchronous bunch of oscillators, depending on the individual natural frequency $\omega_j$. In the case of a uniformly rotating mean field $Z_1$, one can switch to the rotating frame and consider the time-independent probability distribution of $\psi_j=\varphi_j-\arg{Z_1}$. In the rotating frame, Eq.~(\ref{eq113}) for the Fourier coefficients of the PDF of the subpopulation of oscillators with the same natural frequency $\omega$ reads
\begin{equation}
\dot{a}_m=im\delta\omega\,a_m+mh\,a_{m-1}-mh^\ast a_{m+1} +\sigma\dot\Phi^{(\xi)}(m)\, a_m\,,
\label{eq:fe:01}
\end{equation}
where $\delta\omega=\omega-\Omega$, $\Omega$ is the rotation rate of mean field $Z_1$, and $h$ is a frozen constant.

For the derivation of the average oscillation frequency of oscillators with natural frequency $\omega$, one has to deal with the Fokker--Planck equation~(\ref{eq:fe:01}) in the $\psi$-space [cf Eq.~(\ref{eqA4}) in Appendix~\ref{app:FFP}]:
\begin{align}
\frac{\partial w_\omega(\psi,t)}{\partial t}
 +\frac{\partial}{\partial\psi}\Big[\left(\delta\omega -ihe^{-i\psi}+ih^\ast e^{i\psi}\right)w_\omega(\psi,t)\Big]
\nonumber\\
 {}-\sigma\dot\Phi^{(\xi)}\left(i\frac{\partial}{\partial\psi}\right)w_\omega(\psi,t)=0\,.
\label{eq:fe:02}
\end{align}
For a time-independent distribution the probability density flux $J_\omega$, which is defined by the conservation-law form of FPE (\ref{eq:fe:02}), $\frac{\partial}{\partial t}w_\omega+\frac{\partial}{\partial\psi}J_\omega=0$, is constant:
\begin{align}
J_\omega&=\left(\delta\omega -ihe^{-i\psi}+ih^\ast e^{i\psi}\right)w_\omega(\psi)
\nonumber\\
&\qquad {}
-\int\limits^\psi\mathrm{d}\psi_1 \sigma\dot\Phi^{(\xi)}\left(i\frac{\partial}{\partial\psi_1}\right)w_\omega(\psi_1)\,.
\label{eq:fe:03}
\end{align}
The value of the probability density flux gives the average cyclic oscillation frequency detuning $\nu_\omega\equiv\langle\dot\varphi_\omega\rangle-\Omega=\langle\dot{\psi}\rangle=2\pi J_\omega$.
Substituting the Fourier expansion into (\ref{eq:fe:03}), one finds
\begin{align}
J_\omega&=\sum_{m=-\infty}^{+\infty}\left(\delta\omega -ihe^{-i\psi}+ih^\ast e^{i\psi}\right)\frac{a_m}{2\pi}e^{-im\psi}
\nonumber\\
&\quad {}
-\int\limits^\psi\mathrm{d}\psi_1
 \sum_{m=-\infty}^{+\infty}\sigma\left(-|m|+i\frac{2\beta}{\pi}m\ln|m|\right)\frac{a_m}{2\pi}e^{-im\psi_1},
\nonumber
\end{align}
the constant part of which gives
\begin{align}
\nu_\omega=2\pi J_\omega=\delta\omega -\mathrm{Im}(2h^\ast a_1)\,.
\label{eq:fe:04}
\end{align}
For the average oscillation frequency, one calculates the time-independent solution of equation chain~(\ref{eq:fe:01}) and substitute $a_1$ into (\ref{eq:fe:04}).

With the 2CC model reduction~(\ref{eq115})--(\ref{eq116}) (with $Z_1:=a_1$, $\kappa_3=0$ and $\gamma=0$ since we deal with a subpopulation of oscillators with identical natural frequencies), for a fixed field $h$, one expresses
$\kappa_2=\frac{i(2\beta\sigma/\pi)\ln{2}}{\sigma-i\delta\omega+2h^\ast a_1 -i(2\beta\sigma/\pi)\ln{2}}a_1^2$ from (\ref{eq116}), substitutes it into (\ref{eq115}), and finds a cubic equation for $A_1\equiv a_1|h|/h$:
\begin{align}
&A_1^3-i\frac{3}{2}\frac{\delta\omega+i\sigma}{|h|}A_1^2
-\left[1 +\frac{(\delta\omega+i\sigma)(\delta\omega+i\sigma+\frac{2\beta\sigma}{\pi}\ln{2})}{2|h|^2}\right]A_1
\nonumber\\
&
+i\frac{\delta\omega+i\sigma+\frac{2\beta\sigma}{\pi}\ln{2}}{2|h|}=0\,,
\end{align}
the physically relevant solution of which is
\begin{align}
&
a_1=\frac{h}{|h|}\left(i\frac{\delta\omega+i\sigma}{2|h|}+\frac{p^{1/3}}{2} \right.
\nonumber\\
&\qquad\quad
\left.
{}-\frac{(\delta\omega+i\sigma)(\delta\omega+i\sigma-\frac{4\beta\sigma}{\pi}\ln{2})-4|h|^2}{6|h|^2p^{1/3}}
\right)\,,
\label{eq:fe:05}
\\
&p=i\frac{2\beta\sigma\ln{2}}{\pi|h|}\left[\frac{(\delta\omega+i\sigma)^2}{|h|^2}-2\right]
-i\bigg[\frac{64}{27}
\nonumber\\
&{}
-\frac{16(\delta\omega+i\sigma+\frac{(\sqrt{13}-2)\beta\sigma}{\pi}\ln{2}) (\delta\omega+i\sigma-\frac{(\sqrt{13}+2)\beta\sigma}{\pi}\ln{2})}{9|h|^2}
\nonumber\\
&{}
+\frac{4(\delta\omega+i\sigma)^2(\delta\omega+i\sigma+\frac{2\beta\sigma}{\pi}\ln{2}) (\delta\omega+i\sigma-\frac{10\beta\sigma}{\pi}\ln{2})}{9|h|^4}
\nonumber\\
&-\frac{(\delta\omega+i\sigma)^3(\delta\omega+i\sigma+\frac{2\beta\sigma}{\pi}\ln{2})^2 (\delta\omega+i\sigma-\frac{16\beta\sigma}{\pi}\ln{2})}{27|h|^6}\bigg]^\frac12.
\nonumber
\end{align}
(Notice, two other formal solutions of the cubic equation for $A_1$ give either an unstable time-independent state or unphysical solutions with $|a_1|>1$.)
Hence, Eq.~(\ref{eq:fe:04}) within the framework of the 2CC model reduction becomes
\begin{align}
\nu_\omega=|h|\,\mathrm{Im}\left[
\frac{(\delta\omega+i\sigma)(\delta\omega+i\sigma+\frac{4\beta\sigma}{\pi}\ln{2})-4|h|^2}{3|h|^2p^{1/3}}
-p^\frac{1}{3}\right].
\label{eq:fe:05}
\end{align}

For $\beta=0$, Eqs.~(\ref{eq:fe:04})--(\ref{eq:fe:04}) yield exact results as the neglected higher-order CCs tend to zero in this case. Hence, we recover the known solution for a symmetric Cauchy distribution of natural frequencies: 
$a_1=\frac{h}{2|h|^2}\big(i\delta\omega-\sigma+\sqrt{4|h|^2-(\delta\omega+i\sigma)^2}\big)$ and $\nu_\omega
=\mathrm{Im}\sqrt{4|h|^2-(\delta\omega)^2+\sigma^2+2i\delta\omega\,\sigma}$.\footnote{For $\sigma=0$, this solution recovers solutions from Refs.\cite{Klinshov-Franovic-2019,Yasmine-etal-2022}.
}

\begin{figure}[!t]
\center{
\includegraphics[width=0.455\textwidth]%
 {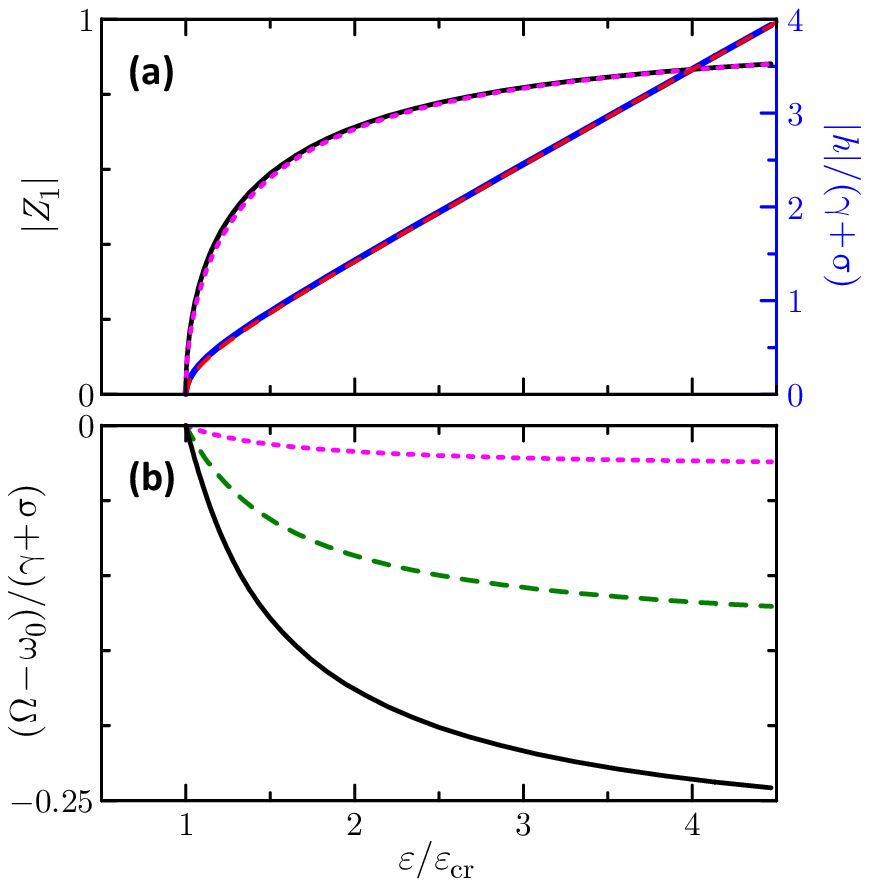}
}
\caption{Macroscopic dynamics of Kuramoto ensemble subject to asymmetric Cauchy noise. (a):~Kuramoto order parameter $|Z_1|$ plotted versus the coupling strength $\varepsilon$ for $\beta=0.1$ (magenta dotted line) and $\beta=1$ (black solid line) is apparently weakly influenced by the noise asymmetry; the dependence of corresponding $|h|$ is plotted for $\beta=1$ (thick blue line; the scale is on the right axis) and well approximated by its asymptotic value (\ref{eq:h_inf}) for $\varepsilon/\varepsilon_\mathrm{cr}\to\infty$ (red dashed line), which is independent of $\beta$.
(b):~The frequency of collective oscillations $\Omega$ is affected by the noise asymmetry: $\beta=0.1$ (magenta dotted), $0.5$ (green dashed), $1$ (black solid).}
  \label{fig4}
\end{figure}

\begin{figure*}[!t]
\centerline{
\includegraphics[width=0.225\textwidth]%
 {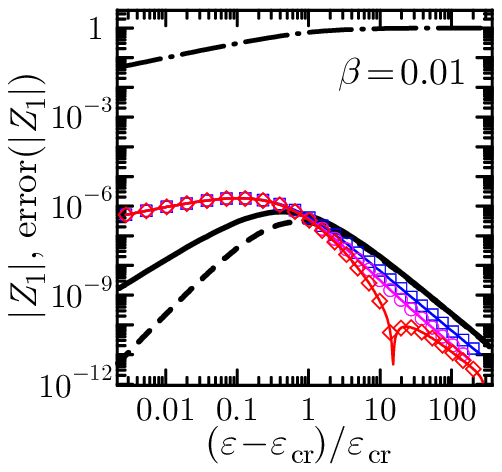}
\quad
\includegraphics[width=0.225\textwidth]%
 {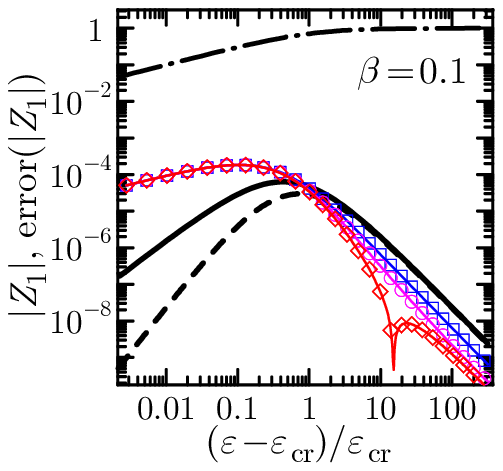}
\quad
\includegraphics[width=0.225\textwidth]%
 {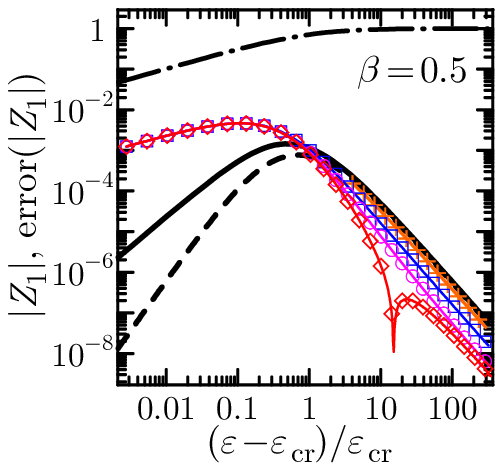}
\quad
\includegraphics[width=0.225\textwidth]%
 {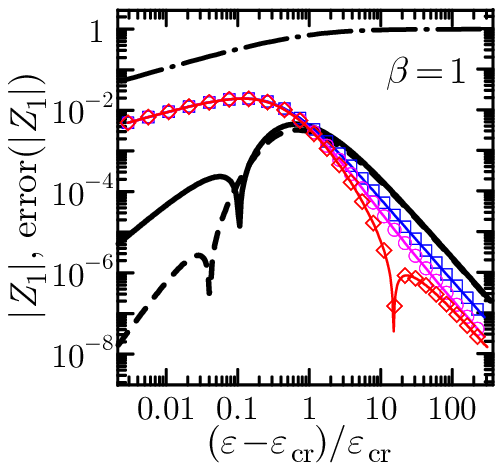}
}

\vspace{10pt}

\centerline{
\includegraphics[width=0.225\textwidth]%
 {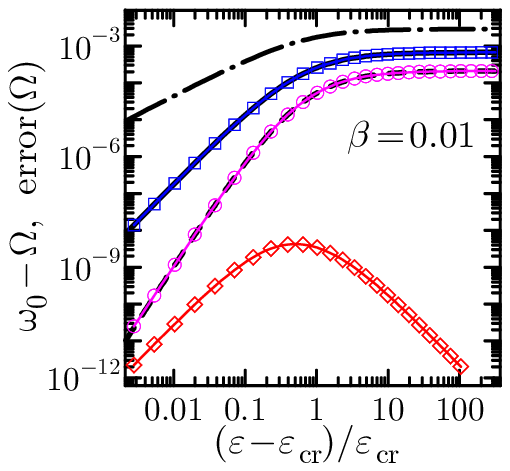}
\quad
\includegraphics[width=0.225\textwidth]%
 {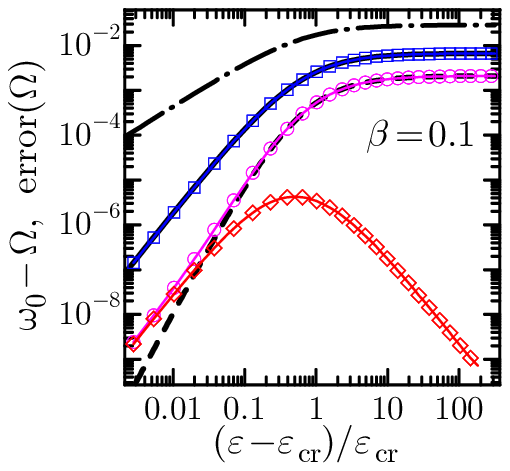}
\quad
\includegraphics[width=0.225\textwidth]%
 {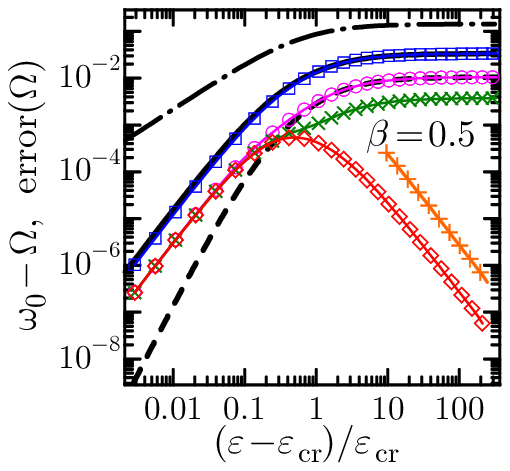}
\quad
\includegraphics[width=0.225\textwidth]%
 {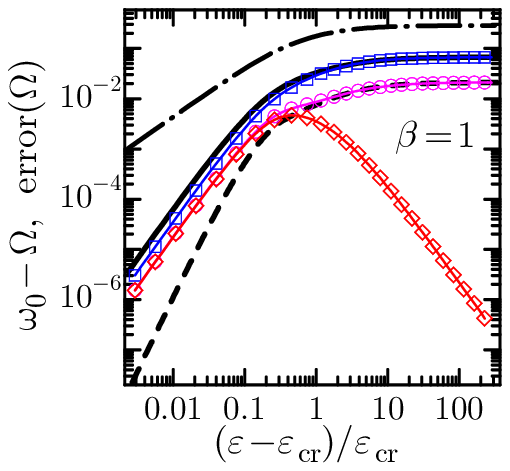}
}
\caption{
The level of synchronization in the Kuramoto ensemble with asymmetric Cauchy noise is given by the ``exact'' value of the Kuramoto order parameter $|Z_1|$ (black dash-dotted line) calculated with the continued fraction method with $m_\infty=2000$ modes [Eqs.~(\ref{eq114_ti_CF}) and (\ref{eq:hbinf})]; the value of noise asymmetry parameter $\beta$ is indicated in plots.
Bold black lines: the error of the 2CC model reduction solution~(\ref{eq:KurZre})--(\ref{eq:KurZim}) (solid line) and of the 3CC model reduction~(\ref{eq115})--(\ref{eq117}) (dashed line);
thin lines with symbols: the error of rigorous linear-in-$\beta\sigma$ solutions (\ref{eq115_ti},\ref{eq:kappa2ti}) truncated after $l=2$ (blue squares), $3$ (magenta circles), $4$ (green crests, shown only for $\beta=0.5$), and $50$ (red diamonds); the plotted segment of the curves for $l=50$ does not change with the increase of $l$ by a factor of two. For $\beta=0.5$, orange pluses: the error of asymptotic analytic solution~(\ref{eq:GZ1infty})--(\ref{eq:GOmegainfty}) for strong coupling.
}
  \label{fig5}
\end{figure*}

\section{Bias of collective oscillation frequency by asymmetric Cauchy noise}\label{sec:bias}
\subsection{Collective dynamics of the Kuramoto ensemble subject to intrinsic asymmetric Cauchy noise}\label{ssec:Kur}

For the Kuramoto ensemble~\cite{Kuramoto-1975,Kuramoto-1984,Acebron-etal-2005} with noise
\begin{equation}
\dot\varphi_j=\omega_j+\frac{\varepsilon}{N}\sum_{m=1}^{N}\sin(\varphi_m-\varphi_j)+\sigma\xi_j(t)\,,
\quad j=1,2,...,N,
\label{eq201}
\end{equation}
where $\varepsilon$ is the coupling strength,
one finds the form~(\ref{eq110a}) with
\[
h(t)=\frac{\varepsilon Z_1}{2}\,.
\]
In the thermodynamic limit $N\to\infty$, one finds a uniformly rotating solution: $\dot{Z}_m=im\Omega Z_m$ or, equivalently, $\dot{\kappa}_m=im\Omega\kappa_m$, where $\widetilde\omega=\Omega-\omega_0$ can be nonzero because of the noise asymmetry $\beta$.
Eq.~(\ref{eq116}) with $\kappa_3=0$ yields
\[
\kappa_2=\frac{i\frac{2\beta\sigma}{\pi}\ln{2}}{\gamma+\sigma+i\widetilde{\omega}+\varepsilon|Z_1|^2 -i\frac{2\beta\sigma}{\pi}\ln{2}}Z_1^2
\]
and, after substitution of $\kappa_2$, Eq.~(\ref{eq115}) can be recast as
\begin{align}
&|Z_1|^4+\left(\frac{3\varepsilon_\mathrm{cr}}{2\varepsilon}-1 +i\frac{3\widetilde{\omega}}{\varepsilon}\right)|Z_1|^2
\nonumber\\
&+\left(\frac{\varepsilon_\mathrm{cr}}{\varepsilon}-1 +i\frac{2\widetilde{\omega}}{\varepsilon}\right)
\left(\frac{\varepsilon_\mathrm{cr}}{2\varepsilon}+i\left[\frac{\widetilde{\omega}}{\varepsilon} -\frac{2\beta\sigma}{\pi\varepsilon}\ln{2}\right]\right)=0\,,
\label{eq:KurZ}
\end{align}
where $\varepsilon_\mathrm{cr}=2(\gamma+\sigma)$ is the Kuramoto transition point, which is unaffected by the noise asymmetry $\beta$; the collective mode exists for $\varepsilon>\varepsilon_\mathrm{cr}$.

For a symmetric noise with $\beta=0$, Eq.~(\ref{eq:KurZ}) is a quadratic equation for $|Z_1|^2$, where $\widetilde\omega=0$. However, for $\beta\ne0$, this equation is more complex than quadratic, because $|Z_1|$ is real-valued, while the coefficients are complex and variable $\widetilde\omega\ne0$ is to be determined from the self-consistency condition. Specifically, the real and imaginary parts of Eq.~(\ref{eq:KurZ}) yield
\begin{align}
&|Z_1|^4+\left(\frac{3\varepsilon_\mathrm{cr}}{2\varepsilon}-1\right)|Z_1|^2
\nonumber\\
&\quad
-\frac{\varepsilon_\mathrm{cr}}{2\varepsilon}\left(1-\frac{\varepsilon_\mathrm{cr}}{\varepsilon}\right)
-\frac{2\widetilde{\omega}}{\varepsilon}
\left(\frac{\widetilde{\omega}}{\varepsilon} -\frac{2\beta\sigma}{\pi\varepsilon}\ln{2}\right)=0\,,
\label{eq:KurZre}
\\
&\frac{\widetilde{\omega}}{\varepsilon}\left(|Z_1|^2
+\frac{2\varepsilon_\mathrm{cr}}{3\varepsilon}-\frac{1}{3}\right)
+\frac{\ln{2}}{3}\frac{2\beta\sigma}{\pi\varepsilon}\frac{\varepsilon_\mathrm{cr}-\varepsilon}{\varepsilon}=0\,.
\label{eq:KurZim}
\end{align}
These two equations determine the values of unknown real-valued variables $(|Z_1|,\widetilde{\omega})$.
%
%
Nontrivial solution to~(\ref{eq:KurZre})--(\ref{eq:KurZim}) can be found to be unique for $\varepsilon>\varepsilon_\mathrm{cr}$; the bifurcation is always supercritical and no solutions below the threshold.


In Fig.~\ref{fig4} one can see the dependence of the Kuramoto order parameter (Fig.~\ref{fig4}a) and the collective frequency bias~(b) on the coupling strength calculated `exactly' by means of the continued fraction method with $2000$ modes $Z_m$ (see Sec.~\ref{sec:contfrac} for detail). For the obtained macroscopic states, the mean field $|h|$, which governs the frequency entrainment of individual oscillators, is also plotted in Fig.~\ref{fig4}a. Here one can see that its value can be estimated with asymptotic law
\begin{equation}
|h|_\infty=\sqrt{\varepsilon^2/4-\varepsilon(\gamma+\sigma)/2}
\label{eq:h_inf}
\end{equation}
valid for $\varepsilon/\varepsilon_\mathrm{cr}\gg1$.

In the top raw of Fig.~\ref{fig5}, one can see that 2- and 3CC model reductions~\footnote{Eqs.~(\ref{eq:KurZre}) and (\ref{eq:KurZim}) are written for the 2CC reduction; for the 3CC reduction analogous equation system is much more cumbersome and we take numerical solution of time-independent version of Eqs.~(\ref{eq115})--(\ref{eq117}) with $\kappa_4=0$.} provide superior accuracy close to the Kuramoto transition point, where the smallness hierarchy takes the form $|\kappa_m|\propto |Z_1|^m$, in agreement to case~(i) discussed at the end of Sec.~\ref{ssec:CCR}. However, for a strong coupling, where we expect to have the hierarchy case~(ii) of Sec.~\ref{ssec:CCR}, not only the 3CC reduction possesses the same order of accuracy as the 2CC reduction but also the accuracy of both becomes inferior to the rigorous asymptotic solution~(\ref{eq:selfconsZ1}), which accounts only for linear in $\beta\sigma$ contributions. The 2- and 3CC reductions yield a decent accuracy uniformly over the entire parameter space. The asymptotic solution~(\ref{eq:selfconsZ1}) is highly accurate for a strong coupling. Its inaccuracy rises for $\beta\sim 1$ and $(\varepsilon-\varepsilon_\mathrm{cr})/\varepsilon_\mathrm{cr}<1$,
where the formal small parameter $\beta\sigma$ of expansion is nonsmall, while a different small parameter emerges: in this case, $|Z_1|\ll 1$.
Nonetheless, even in this domain of the parameter space, the accuracy is reasonably good.

\subsection{Nonlinear noise-induced bias of the mean field rotation}
In Fig.~\ref{fig4}b and the bottom row of Fig.~\ref{fig5}, one can see the bias $\widetilde\omega=\Omega-\omega_0$ of the rotation of the mean field; this bias is induced by the noise asymmetry and emerges as a nonlinear effect: it depends on the amplitude of the collective mode and vanishes close to the Kuramoto transition point, where the collective mode amplitude tends to zero. From the equation chain~(\ref{eq114}) one can clearly see that for a given Fourier mode $Z_m$ the $\beta$-term enforces a faster rotation $\propto\beta\sigma\ln|m|$; the effect of the noise asymmetry on the rotation of the mean field is stronger for higher $m$, that is for the narrower distributions.

\begin{figure*}[!t]
\centerline{
\includegraphics[width=0.310\textwidth]%
 {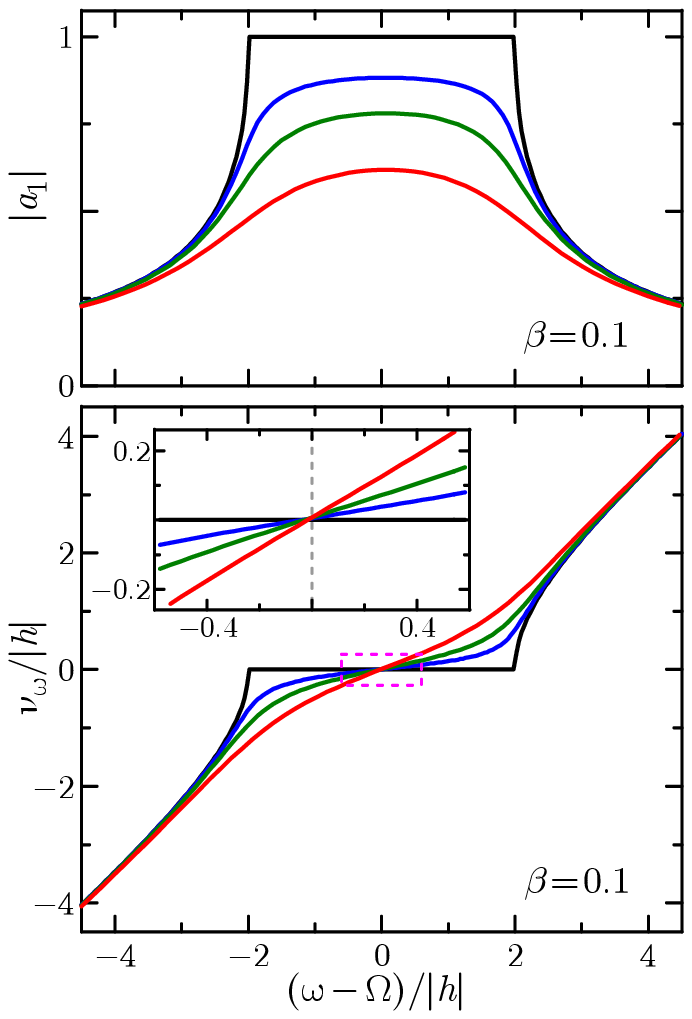}
\quad
\includegraphics[width=0.310\textwidth]%
 {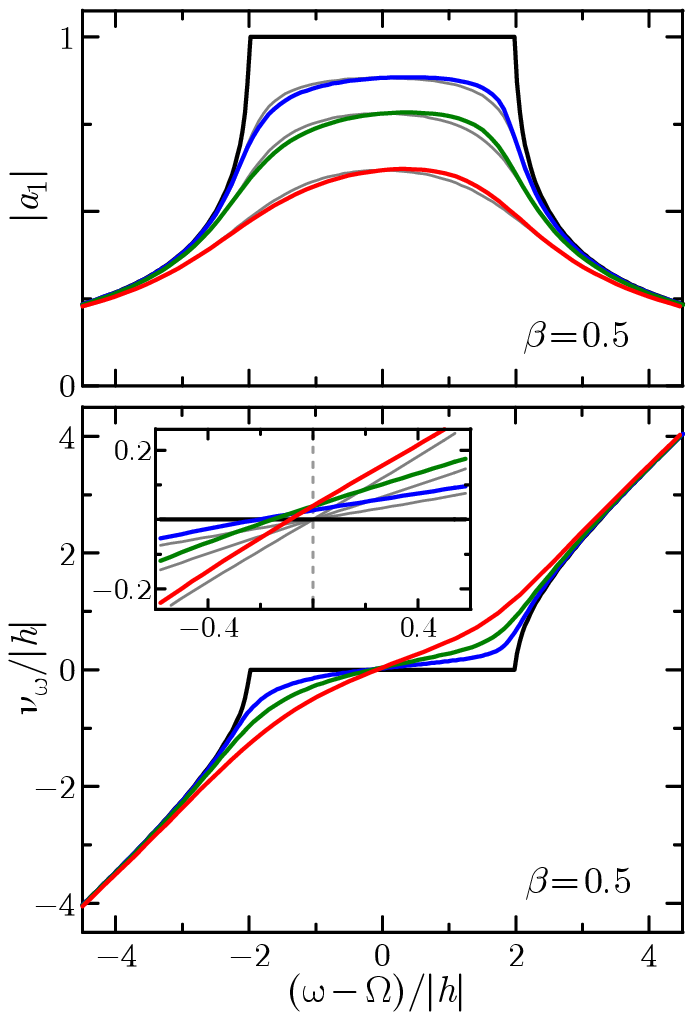}
\quad
\includegraphics[width=0.310\textwidth]%
 {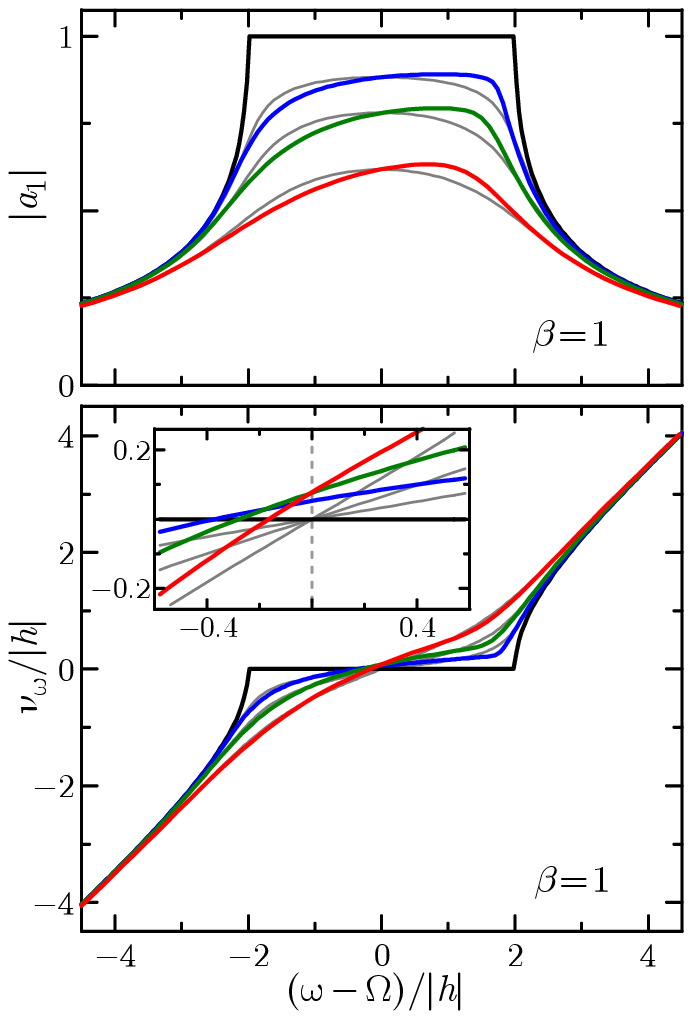}
}

\caption{The symmetry of the effects of entrainment of frequencies (lower panels) and subpopulations (upper panels) of oscillators with natural frequency $\omega$ is broken for asymmetric noise. From top to bottom in the upper panels $\sigma/|h|=0$, $0.25$, $0.5$, and $1$; the color coding is the same in all panels; the magenta dotted rectangle marks the graph area zoomed in insets.
For the reference, the thin gray lines represent the dependencies for a symmetric noise ($\beta=0$); not shown for the cases where their deviation from the color lines is indistinguishable against the background of the line thickness.
}
  \label{fig6}
\end{figure*}

One can derive the asymptotic law of $\widetilde\omega(\varepsilon)$ for a strong coupling $\varepsilon\gg\varepsilon_\mathrm{cr}$. For a strong coupling the 2CC model reduction provides decent accuracy for the absolute value of the order parameter $Z_1$; however, the inaccuracy of the rotation bias is small but does not tend to zero in the limit $\varepsilon/\varepsilon_\mathrm{cr}\to\infty$ (Fig.~\ref{fig5}). Meanwhile the rigorous expansion~(\ref{eq115_ti})--(\ref{eq116_ti}), by construction, accounts for the fact that the higher CCs are small compared to $|Z_1|$ but their decay relatively each other can be not as fast; mathematically, $|\kappa_2/Z_1|\ll|\kappa_{m+1}/\kappa_{m}|<1$. In Fig.~\ref{fig5} with the red dotted lines, one can see the inaccuracy of the rotation bias $\widetilde\omega$ given by (\ref{eq115_ti})--(\ref{eq116_ti}) tends to zero for a stronger coupling as the number of terms kept in the series (\ref{eq:kappa2ti}) grows. For $\varepsilon\gg\varepsilon_\mathrm{cr}$, one can use the smallness of $(\gamma+\sigma+i\widetilde\omega)/\varepsilon$ and $(1-|Z_1|)$ to simplify self-consistency equation~(\ref{eq:selfconsZ1}):
\begin{align}
&\gamma+\sigma+i\widetilde\omega-\frac{\varepsilon}{2}(1-|Z_1|^2)=
-i\frac{2\beta\sigma}{\pi}\bigg[S_0-S_1\frac{\gamma+\sigma+i\widetilde\omega}{\varepsilon}
\nonumber\\
&{}+\mathcal{O}_1\left(\frac{(\gamma+\sigma+i\widetilde\omega)^2}{\varepsilon^2}\right)
+\mathcal{O}_2\left(\frac{\gamma+\sigma+i\widetilde\omega}{\varepsilon}(1-|Z_1|^2)\right)
\bigg]
\,,
\label{eq:selfconsZ1app}
\\
&\qquad
S_n=\sum_{l=1}^\infty\frac{(-1)^{l+1}l^n l!}{2^l}C_{l+1}\,,
\nonumber
\end{align}
where $\mathcal{O}_j(\cdot)$ stand for functions of the order of magnitude of their small arguments,
$$
S_0=0.451582705...\,,
\quad 
S_1=0.609274778...\,.
$$ 
Solving the real and imaginary parts of Eq.~(\ref{eq:selfconsZ1app}), one obtains
\begin{align}
|Z_1|^2&=1-2\frac{\gamma+\sigma}{\varepsilon}+\frac{8S_0S_1\beta^2\sigma^2}{\pi^2\varepsilon^2}
+\dots\,,
\label{eq:GZ1infty}
\\
\widetilde\omega&=\frac{2\beta\sigma}{\pi}\left(-S_0+S_1\frac{\gamma+\sigma}{\varepsilon}\right)
+\dots\,,
\label{eq:GOmegainfty}
\end{align}
where ``$\dots$'' stands for higher order corrections.

In Fig.~\ref{fig5}, for the effect which is completely discarded within the framework of the OA Ansatz, the `plain' 2- and 3CC model reductions are useful to grasp the effect and give converging results where the effect is small, but yield a finite non-decaying error for the asymptotic behavior for a strong coupling, where the effect is nonsmall. The rigorous expansion~(\ref{eq115_ti},\ref{eq:kappa2ti}) provides uniformly accurate theory of the collective frequency bias effect; the error of the explicit asymptotic ($\varepsilon/\varepsilon_\mathrm{cr}\to\infty$) solution~(\ref{eq:GZ1infty})--(\ref{eq:GOmegainfty}) for this expansion is shown with orange pluses for $\beta=0.5$ in Fig.~\ref{fig5}.

\subsection{Individual frequency entrainment}
\label{ssec:entrainKur}
In Fig.~\ref{fig6}, the individual oscillator frequencies are plotted versus their natural frequency for the noisy Kuramoto ensemble. The value of the mean field $|h|$ is provided in Fig.~\ref{fig4}a as a function of the ensemble parameters. In the noise-free case, one observes a perfect synchronization $|a_1|=1$ of the subpopulations with $|\omega-\Omega|<2|h|$ to the rotating mean field. Outside this natural frequency band one observes a grouping of states ($|a_1|>0$) and imperfect frequency entrainment which decreases at distance from the synchronization frequency band. In the presence of noise, the synchronization and frequency locking within the band $|\omega-\Omega|<2|h|$ are never perfect. One can notice a pronounced asymmetry owned by the noise asymmetry: for $\beta>0$, the subpopulations with fast natural frequencies ($\beta(\omega-\Omega)>0$) are more coherent (bigger values of $|a_1|$). The average oscillation frequencies $\nu_\omega$ are asymmetric, but the sign and magnitude of this asymmetry depend nonmonotonously on $|\omega-\Omega|$; for different absolute values of the detuning, faster or slower oscillators can exhibit smaller deviation of the individual average frequency from the mean field frequency.

The presented results are calculated with the continued fraction method (see Sec.~\ref{sec:contfrac}). The results calculated with the 2CC model reduction (Sec.~\ref{sec:entrain}) are visually undistinguishable from the graphs in Fig.~\ref{fig6}.
One can also see that the asymmetry of the frequency entrainment effect produced by the noise asymmetry is less pronounced than the effect of the nonlinear bias of collective oscillation frequency (Fig.~\ref{fig4}b); the former is noticeable only for $\beta\sim1$ (compare gray and color lines in the bottom row of Fig.~\ref{fig6}).

\subsection{Continued fraction solution for time-independent states}\label{sec:contfrac}
High precision numerical solution for time-independent states can be obtained by means of the continued fraction method,~\cite{Khinchin-1964} which we recall here in application to our problem.
The time-independent form of equation system~(\ref{eq114}) for $\dot{Z}_m=im\Omega Z_m$ reads
\begin{equation}
0=-(i\widetilde{\omega}+\gamma+\sigma)Z_m+hZ_{m-1}-h^\ast Z_{m+1}
 +i\frac{2\beta\sigma}{\pi}\ln{m}\, Z_m\,.
\label{eq114_ti}
\end{equation}
For $B_m=Z_m/Z_{m-1}$, this equation system takes the form of
\begin{equation}
0=i\left(\frac{2\beta\sigma}{\pi}\ln{m}-\widetilde{\omega}\right)-\gamma-\sigma
 +\frac{h}{B_m}-h^\ast B_{m+1}\,.
\label{eq114_ti_b}
\end{equation}
Equivalently, $B_m=\frac{h}{\gamma+\sigma+i\left(\widetilde{\omega}-\frac{2\beta\sigma}{\pi}\ln{m}\right)
+h^\ast B_{m+1}}$. As $Z_0=1$, one can write for $Z_1=B_1$ a continuous-fraction
\begin{widetext}
\begin{equation}
Z_1=\frac{h}{\displaystyle \gamma+\sigma+i\widetilde{\omega}+
\frac{|h|^2}{\displaystyle \gamma+\sigma+i\left(\widetilde{\omega}-\frac{2\beta\sigma}{\pi}\ln{2}\right)+
\frac{|h|^2}{\displaystyle \gamma+\sigma+i\left(\widetilde{\omega}-\frac{2\beta\sigma}{\pi}\ln{3}\right)+
\frac{|h|^2}{\displaystyle \gamma+\sigma+i\left(\widetilde{\omega}-\frac{2\beta\sigma}{\pi}\ln{4}\right)+
\frac{|h|^2}{\dots}
}}}}\,.
\label{eq114_ti_CF}
\end{equation}
\end{widetext}
For practical calculations we need to make truncation at some large $m_\infty\gg1$; for which we adopt the ``self-similarity'' approximation $B_{m_\infty+1}=B_{m_\infty}$:
\[
h^\ast B_{m_\infty}=\frac{|h|^2}{\gamma+\sigma +i\left(\widetilde{\omega}-\frac{2\beta\sigma}{\pi}\ln{m_\infty}\right)
+h^\ast B_{m_\infty}}\,.
\]
The latter equation is solved with
\begin{align}
h^\ast B_{m_\infty}&=-\frac{1}{2}
\left(\gamma+\sigma+i\left[\widetilde{\omega}-\frac{2\beta\sigma}{\pi}\ln{m_\infty}\right]
\right)\times
\nonumber\\
&\left(1\pm\sqrt{1+\frac{4|h|^2}{\left(\gamma+\sigma +i\left[\widetilde{\omega}-\frac{2\beta\sigma}{\pi}\ln{m_\infty}\right]
\right)^2}}\right)\,,
\label{eq:hbinf}
\end{align}
where the ``minus''-solution gives $B_{m_\infty}\approx2h/\{\gamma+\sigma +i\left[\widetilde{\omega}-(2\beta\sigma/\pi)\ln{m_\infty}\right]\}$ for large $m_\infty$ and the series converges, while the ``plus''-solution diverges; therefore, the ``minus''-branch of~(\ref{eq:hbinf}) is of physical interest.

For the Kuramoto system, one can calculate the ``exact'' solution in a parametric form as follows. For given $(\gamma,\sigma,\beta\sigma)$, one fixes $|h|^2$, starting from (\ref{eq:hbinf}) and using (\ref{eq114_ti_CF}) calculates $(h^\ast Z_1)$ as a function of $\widetilde{\omega}$, finds unknown $\widetilde{\omega}$ providing $\mathrm{Im}(h^\ast Z_1)=0$. Therefore, for a fixed value of $|h|^2=\varepsilon^2|Z_1|^2/4$, one finds the value of $h^\ast Z_1=\varepsilon|Z_1|^2/2$; hence, $\varepsilon=2|h|^2/(h^\ast Z_1)$ and $|Z_1|^2=(h^\ast Z_1)^2/|h|^2$.

For the frequency entrainment effect considered in Sec.~\ref{sec:entrain}, the calculations via continued fractions are even more simple: $Z_1$ to be replaced with $a_1$, $\gamma=0$, and the value of $\widetilde{\omega}=-\delta\omega$ is given.


\section{Conclusion}\label{sec:concl}
We have derived finite (low) dimensional model reductions for macroscopic dynamics of thermodynamically large populations of sin-coupled phase oscillators subject to asymmetric Cauchy noise [Eqs.~(\ref{eq115})--(\ref{eq117})]. In contrast to the case of symmetric Cauchy (Lorentzian) noise, the asymmetry enforces deviations from the Ott--Antonsen manifold $Z_m=(Z_1)^m$. Here one can employ the circular cumulant (CC) representation, which typically decays fast enough to capture the macroscopic dynamics with just two or three first CCs.

Within the framework of the CC model reductions we revealed and studied the effect of the bias of the collective oscillation frequency owned by the noise asymmetry. The effect was quantitatively characterised in detail for the Kuramoto ensemble [Eqs.~(\ref{eq:KurZre}) and (\ref{eq:KurZim}), Fig.~\ref{fig4}]. The bias magnitude depends on the amplitude of the collective mode and vanishes next to the Kuramoto transition point; the effect is essentially nonlinear. The asymptotic law for the case of a strong coupling was derived analytically [Eqs.~(\ref{eq:GZ1infty}) and (\ref{eq:GOmegainfty})].

The frequency bias effect is shaped by two factors: skewness and heavy tails of noise. The role of these factors cannot be disentangled, since the asymmetry essentially forces the bias, but this asymmetry does not exist without heavy tails. Indeed, by virtue of the Generalized central limit theorem, microscopic fluctuations without heavy tails effectively generate a Gaussian noise, which cannot be skewed.
The theoretical study for heavy-tailed noises necessitated an upgrade of mathematical framework, which was delivered on the basis of the CC approach.\cite{Tyulkina-etal-2018,Goldobin-Dolmatova-2019b,Dolmatova-Tyulkina-Goldobin-2023}

For populations of heterogeneous oscillators subject to asymmetric Cauchy noise, the theoretical description of the individual oscillator frequency entrainment by the collective mode was constructed within the framework of the finite CC model reductions [Eq.~(\ref{eq:fe:05})]. The detailed quantitative characterisation of the effect was provided for the Kuramoto ensemble (Fig.~\ref{fig6}).

Validation of the theoretical results against the background of the ``exact'' numerical solutions revealed decent accuracy of the finite CC model reductions uniformly over the entire parameter space (Fig.~\ref{fig5}). The ``exact'' results were calculated by means of the continued fraction method with $2000$ Fourier modes $Z_m$ and nearly machine accuracy.

While the Gaussian distribution is symmetric by its nature, the non-Gaussian stable distributions are often asymmetric and the emergence of this asymmetry is robust, which makes the case of asymmetric Cauchy noise physically relevant in many cases, where microscopic fluctuations possess a Cauchy-type heavy tails.

The symmetric $\alpha$-stable fluctuations are still not a mathematically degenerate case but must be quite generic under various realistic circumstances. For instance, in a population of globally (linearly) coupled heterogeneous oscillators synchronized by common Gaussian noise,\cite{Goldobin-Dolmatova-2019} the state deviations possess symmetric power-law tails with $\alpha=1+\varepsilon/|\varepsilon_\ast|$, where $-|\varepsilon_\ast|$ is the threshold coupling strength (negative) which is needed to prevent synchronization by common noise. On a long-term scale, these state deviations generate a stable effective noise with $\alpha=1+\varepsilon/|\varepsilon_\ast|$ for $-|\varepsilon_\ast|<\varepsilon<|\varepsilon_\ast|$ and Gaussian noise for a strong attractive coupling $\varepsilon\ge|\varepsilon_\ast|$. These effective noises are symmetric, without any symmetries in original coupled oscillators, which are general limit-cycle oscillators with a general form of the phase resetting curve.\cite{Goldobin-Dolmatova-2019}
The synchronization of coupled oscillators by common noise is
a theoretical framework for a response of coupled populations to an irregular driving signal. Therefore, the case of self-organized endogenous $\alpha$-stable noise without asymmetry should be still quite abundant for some types of couplings. Meanwhile, as we explained in Sec.~\ref{sec:ASN}, pulse-type coupling networks (e.g., synaptic) are expected to generate asymmetric effective white noise as soon as this noise is non-Gaussian.

Presumably, as an implication, one could expect for neuronal populations with sparse random networks with the electrical gap and synaptic couplings,\cite{Montbrio-Pazo-2020} the gap couplings to generate a symmetric effective endogenous noise and the synaptic couplings to generate an asymmetric one.
For neuronal populations the case of asymmetric Cauchy noise can be important. The impact of a symmetric Cauchy noise within the framework of the ``next-generation neural mass models'' is equivalent to the impact of the Lorentzian quenched disorder and does not give room for the collective noise-induced oscillations. On the contrary, in the presence of asymmetry, away from the Ott--Antonsen manifold $Z_m=(Z_1)^m$, the occurrence of collective oscillations owned by asymmetric Cauchy noise might be not excluded.

We presented the theoretical and ``exact'' numerical results for the asymmetric Cauchy noise which is a special case $\alpha=1$ of L\'evy noise.  The case $\alpha\ne1$ requires separate treatment as the skewness parameter $\beta$ creates different coefficients of the $|k|^\alpha$-term in the characteristic function of noise (\ref{eq101}) for positive and negative $k$ for $\alpha\ne1$, while it complements the $|k|$-term with $ik\ln|k|$ for $\alpha=1$---the addition has a different shape [see Eq.~(\ref{eq102})].
The former case can be studied in the same spirit as the case $\alpha=1$ but all derivations will be different.

\section*{ACKNOWLEDGMENTS}
The work was carried out as part of a major scientific project (Agreement no. 075-15-2024-535 by April 23, 2024).

\section*{AUTHOR DECLARATIONS}
\subsection*{Conflict of Interest}
The authors have no conflicts to disclose.

\subsection*{Author Contributions}
\textbf{Maria V.\ Ageeva:}
Conceptualization (supporting);
Formal analysis (equal);
Investigation (equal);
Software (equal);
Writing -- original draft (supporting).
\textbf{Denis S.\ Goldobin:}
Conceptualization (lead);
Formal analysis (equal);
Investigation (equal);
Software (equal);
Writing -- original draft (lead).

\section*{DATA AVAILABILITY}
The data that support the findings of this study are available within the article and shown in the figures.

\appendix
\section{Fractional Fokker--Planck equation for additive asymmetric Cauchy noise}\label{app:FFP}
In this section we provide a brief derivation of the evolution equation for the probability density function $w(\varphi,t)$ for the stochastic system~(\ref{eq110a}) for continuous time as a limiting case $\Delta t\to0$ of the discrete time with time step $\Delta t$.\cite{Gardiner-1997,Goldobin-Permyakova-Klimenko-2024,Toenjes-Pikovsky-2020} For additive noise, one can evaluate the increment
 $\Delta w(\varphi,t)=w(\varphi,t+\Delta t)-w(\varphi,t)$
for an infinitesimal $\Delta t$:
\begin{align}
w(\varphi,t+\Delta t)&=-\frac{\partial}{\partial\varphi}\Big([\omega(t)-ih(t)e^{-i\varphi}+ih^\ast(t)e^{i\varphi}]w(\varphi,t)\Big)\Delta{t}
\nonumber\\
&{}+\int\limits_{-\infty}^{+\infty}\mathrm{d}\xi_{\Delta t}P(\xi_{\Delta t})\,w(\varphi-\sigma\xi_{\Delta t}\Delta t,t)\;,
\label{eqA1}
\end{align}
where $\xi_{\Delta t}$ is a discrete time version of noise with the time step $\Delta t$ and $P(\xi_{\Delta t})$ is the probability density of $\xi_{\Delta t}$.
In the r.h.s.-part of (\ref{eqA1}), the last term represents the arrival of states from $w(\varphi-\sigma\xi_{\Delta t}\Delta t,t)$ to $w(\varphi,t+\Delta t)$ after noise kick $(\sigma\xi_{\Delta t}\Delta t)$ averaged over the distribution of these kicks; and the first term represents a deterministic flow of probability.
With Fourier series~(\ref{eq:wFour}),
one can rewrite Eq.~(\ref{eqA1}) as
\begin{align}
&w(\varphi,t+\Delta t) +\frac{\partial}{\partial\varphi}\Big([\omega(t)-ih(t)e^{-i\varphi}+ih^\ast(t)e^{i\varphi}]w(\varphi,t)\Big)\Delta{t}
\nonumber\\
&\quad
{}=\sum_{m=-\infty}^{+\infty}
\int\limits_{-\infty}^{+\infty}\mathrm{d}\xi_{\Delta t}P(\xi_{\Delta t})
\frac{a_m(t)}{2\pi}\,e^{-im\left(\varphi-\sigma\xi_{\Delta t}\Delta t\right)}
\nonumber\\
&=w(\varphi,t)
+\sum_{m=-\infty}^{+\infty}
\left[F_{\xi_{\Delta t}}(m;c=\sigma\Delta tc_{\xi_{\Delta t}})-1\right]
\frac{a_m(t)}{2\pi}\,e^{-im\varphi}\,,
\nonumber
\end{align}
where $F_{\xi_{\Delta t}}(m;c)$ stands for the characteristic function (\ref{eq101}) of an $\alpha$-stable variable with scale $c$.
For $\Delta t\to 0$, $F_{\xi_{\Delta t}}(m;\sigma\Delta tc_{\xi_{\Delta t}})
=\exp[\Phi_{\xi_{\Delta t}}(m;\sigma\Delta tc_{\xi_{\Delta t}})]=1+\Phi_{\xi_{\Delta t}}(m;\sigma\Delta tc_{\xi_{\Delta t}})
+[\Phi_{\xi_{\Delta t}}(m;\sigma\Delta tc_{\xi_{\Delta t}})]^2/2+\dots$\,.
Hence,
\begin{align}
&\frac{\Delta w(\varphi,t)}{\Delta t}
 +\frac{\partial}{\partial\varphi}\Big([\omega(t)
 -ih(t)e^{-i\varphi}+ih^\ast(t)e^{i\varphi}]w(\varphi,t)\Big)
\nonumber\\
&\qquad
{}=\sum_{m=-\infty}^{+\infty}
\frac{\Phi_{\xi_{\Delta t}}(m;\sigma\Delta tc_{\xi_{\Delta t}})}{\Delta t}
\frac{a_m(t)}{2\pi}\,e^{-im\varphi}\,.
\label{eqA2}
\end{align}
With $\xi_{\Delta t}\propto(\Delta t)^{1/\alpha-1}$, as discussed in Sec.\ \ref{sec:ASN},
one should take $c_{\xi_{\Delta t}}=(\Delta t)^{1/\alpha-1}$ and find
\begin{equation}
\frac{\Phi_{\xi_{\Delta t}}(m;\sigma\Delta tc_{\xi_{\Delta t}})}{\Delta t}
=\sigma^\alpha\frac{\Phi_{\xi_{\Delta t}}\!\!\left(m;(\Delta t)^{1/\alpha}\right)}{\Delta t}
=:\sigma^\alpha\dot\Phi^{(\xi)}(m)\,.
\label{eqA3}
\end{equation}

Notice here a sophisticated form of the fractional diffusion term in the $\varphi$-space. Eq.~(\ref{eqA2}) can be also formally rewritten as
\begin{align}
&\frac{\partial w(\varphi,t)}{\partial t}
 +\frac{\partial}{\partial\varphi}\Big([\omega(t)
 -ih(t)e^{-i\varphi}+ih^\ast(t)e^{i\varphi}]w(\varphi,t)\Big)
\nonumber\\
&\qquad\qquad\qquad
{}=\sigma^\alpha\dot\Phi^{(\xi)}\left(i\frac{\partial}{\partial\varphi}\right)w(\varphi,t)
\,,
\label{eqA4}
\end{align}
where operator $\dot\Phi^{(\xi)}(i\frac{\partial}{\partial\varphi})$ is nonlocal for $\alpha\ne2$, but has a simple shape in the Fourier space. For the Fourier form of Eq.~(\ref{eqA2}) one can substitute series~(\ref{eq:wFour}) and collect terms $\propto e^{-im\varphi}$ to find
\[
\dot{a}_m=m\left[i\omega(t)+h(t)a_{m-1}-h^\ast(t)a_{m+1}\right]+\sigma^\alpha\dot\Phi^{(\xi)}(m)a_m\,;
\]
for $\alpha=1$ (Cauchy noise) this equation chain is identical to (\ref{eq113}).

The last term of Eq.~(\ref{eqA2}) also corroborates that the discrete time version of noise $\xi_{\Delta t}(t)$ must scale as $\propto(\Delta t)^{1/\alpha-1}$; otherwise, the macroscopic impact of noise, given by the ratio $\Phi_{\xi_{\Delta t}}(m;\sigma\Delta tc_{\xi_{\Delta t}})/(\Delta t)$ (\ref{eqA3}), either vanishes or diverges in the continuous time limit $\Delta t\to0$.







\end{document}